\documentclass[%
 aip,
 amsmath,amssymb,
reprint,%
]{revtex4-1}

\usepackage{graphicx}
\usepackage{caption}
\usepackage{subcaption}

\usepackage{dcolumn}
\usepackage{bm}

\usepackage[utf8]{inputenc}
\usepackage[T1]{fontenc}
\usepackage{mathptmx}
\usepackage{etoolbox}

\makeatletter
\def\@email#1#2{%
 \endgroup
 \patchcmd{\titleblock@produce}
  {\frontmatter@RRAPformat}
  {\frontmatter@RRAPformat{\produce@RRAP{*#1\href{mailto:#2}{#2}}}\frontmatter@RRAPformat}
  {}{}
}%
\makeatother
\begin{document}

\preprint{AIP/123-QED}

\title{Novel 3D Reciprocal Space Visualization of Strain Relaxation in InSb on GaAs Substrates}


\author{T. Blaikie}
\affiliation{ 
Electrical and Computer Engineering, University of Waterloo, 200 University Ave. W, Waterloo, ON N2L 3G1, Canada
}%

\author{Y. Shi}
\affiliation{ 
Physics and Astronomy, University of Waterloo, 200 University Ave. W, Waterloo, ON N2L 3G1, Canada
}%

\author{M. C. Tam}
\affiliation{ 
Electrical and Computer Engineering, University of Waterloo, 200 University Ave. W, Waterloo, ON N2L 3G1, Canada
}%



\author{B. D. Moreno}
\affiliation{ 
BXDS-IVU Beamline, Canadian Light Source Inc., 44 Innovation Blvd., SK S7N 2V3, Canada
}%

\author{Z. R. Wasilewski}
\affiliation{ 
Electrical and Computer Engineering, University of Waterloo, 200 University Ave. W, Waterloo, ON N2L 3G1, Canada
}%
\affiliation{ 
Physics and Astronomy, University of Waterloo, 200 University Ave. W, Waterloo, ON N2L 3G1, Canada
}%
\affiliation{Waterloo Institute for Nanotechnology, 200 University Ave. W, Waterloo, ON N2L 3G1, Canada}


\date{\today}

\begin{abstract}

This study introduces the Reciprocal Space Polar Visualization (RSPV) method, a novel approach for visualizing X-ray diffraction-based reciprocal space data. RSPV allows for the precise separation of tilt and strain, facilitating their individual analysis. InSb was grown by molecular beam epitaxy (MBE) on two (001) GaAs substrates --- one with no misorientation (Sample A) --- one with 2° surface misorientation from the (001) planes (Sample B). There is a substantial lattice mismatch with the substrate and this results in the generation of defects within the InSb layer during growth. To demonstrate RSPV's effectiveness, a comprehensive comparison of surface morphology, dislocation density, strain, and tilt was conducted. RSPV revealed previously unobserved features of the (004) InSb Bragg peak, partially explained by the presence of threading dislocations and oriented abrupt steps (OASs). Surface morphologies examined by an atomic force microscope (AFM) revealed that Sample B had significantly lower root mean square (RMS) roughness. Independent estimates of threading dislocation density (TDD) using X-ray diffraction (XRD) and electron channelling contrast imaging (ECCI) confirmed that Sample B exhibited a significantly lower TDD than Sample A. XRD methods further revealed unequal amounts of $\alpha$ and $\beta$ type threading dislocations in both samples, contributing to an anisotropic Bragg peak. RSPV is shown to be a robust method for exploring 3D reciprocal space in any crystal, demonstrating that growing InSb on misoriented GaAs produced a higher-quality crystal compared to an on-orientation substrate.

\end{abstract}


\maketitle

\section{\label{sec:intro}Introduction}

There has been an increase in the interest of growing single-crystal epitaxial layers of InSb in recent years. InSb is getting more attention for its use in developing plasmonic devices that operate at terahertz (THz) frequencies. 
For plasmonic and other optoelectronic applications InSb should, ideally, be grown on a lattice-matched and semi-insulating substrate. However, there are no such substrates currently available for InSb. Hence, GaAs is frequently chosen as a substrate for InSb, as semi-insulating GaAs wafers are easily obtainable at a relatively low cost.

However, the lattice constant of InSb exceeds that of GaAs by 14.6\%. This substantial mismatch results in the evolution of defects within the InSb layer during growth, serving as a mechanism to alleviate the significant strain. Various strategies have been employed to mitigate threading dislocations and improve surface morphology when growing InSb on GaAs. These approaches include the use of substrates with intentional surface misorientation, initial growth of a low-temperature InSb seed layer, incorporation of an interfacial misfit dislocation array, optimization studies of growth parameters, inclusion of strain-relaxing AlInSb buffer layers, or growing on a (111) substrate surface orientation.\cite{RN412, RN413, RN414, RN415, RN247, RN262, RN275}

In this study, we employ the first two methods to grow InSb on two GaAs (001) substrates, distinguished by their surface orientation. Wafer A aligns with the (001) planes (on-orientation), while Wafer B is intentionally misoriented from the (001) planes. The misorientation of Wafer B entails a surface normal vector tilted 2° towards the [010] crystallographic direction. The study focuses on strain relaxation, with threading dislocation density (TDD) serving as a metric for assessing crystalline quality.

The aim of this research was twofold. Firstly, a study was conducted to analyze the strain and tilt of crystallographic planes caused by significant amounts of dislocations present within two InSb samples. The research explored how substrate surface misorientation affected the surface morphology, dislocation density, strain, and tilt of the grown crystal.

Secondly, this research introduced an innovative technique for exploring 3D reciprocal space, referred to as Reciprocal Space Polar Visualization (RSPV). RSPV was shown to be capable of independently exploring the distributions of strain or tilt within a crystal. Moreover, this technique revealed previously unobserved features about the reciprocal space form of the InSb crystals. Therefore, it provided additional insights into the dislocations present in the samples that conventional X-ray diffraction (XRD) techniques did not provide.

\section{\label{sec:MBE}Growth Procedure}

InSb was deposited on two 3” GaAs (001) wafers in a Veeco Gen10 Molecular Beam Epitaxy (MBE) system utilizing solid-source effusion cells. The As and Sb cells function as cracker cells, with cracking zone temperatures calibrated to generate As$_2$ and Sb$_2$ fluxes. The substrate temperatures reported herein were consistently measured using band-edge thermometry.

The pivotal distinction between the two wafers lay in whether the substrate surface was on-orientation or misoriented. The layer structures grown on the substrates are detailed in Table \ref{tab:Schematic}. Furthermore, an AlAs layer was introduced as a lift-off layer, enabling subsequent removal of the InSb from the substrate if deemed necessary.

\begin{table}
\caption{The layer structure for MBE growths of Sample A and Sample B.}
\label{tab:Schematic}
\begin{ruledtabular}
\begin{tabular}{ccc}
Description			& Material			& Thickness (nm) 	\\
\hline
Substrate			& SI-GaAs (100)		& 6.25E05			\\
Buffer Layer		& GaAs				& 200				\\
Lift-off Layer		& AlAs				& 10				\\
LT InSb Layer		& InSb				& 20				\\
HT InSb Layer		& InSb				& 1500				\\
\end{tabular}
\end{ruledtabular}
\end{table}

The growth procedure unfolded in the following sequence. Initially, substrate oxides were eliminated through annealing at 630 °C under As overpressure. Next, a 200 nm GaAs buffer layer was grown at a rate of 2.0 Å/s, maintaining a substrate temperature of 600 °C. The third step involved growing a 10 nm layer of AlAs at the same temperature but with a rate of 1.0 Å/s. Following this, the substrate was cooled to 400 °C, and the As flux was interrupted while the substrate continued to cool to 200 °C. A low-temperature (LT) layer of InSb, 20 nm thick, was grown at a rate of 2.0 Å/s. Subsequently, all sources had their shutters closed, and the substrate was heated to 350 °C over approximately 7 minutes. Finally, at 2.0 Å/s the high-temperature (HT) InSb layer was deposited with a thickness of 1500 nm. This two-stage growth technique for growing InSb on GaAs was adapted from Refs. \onlinecite{RN275, RN247}.

Reflection high-energy electron diffraction (RHEED) was employed to monitor the growth at the commencement of each new layer. After substrate annealing, the GaAs buffer layer was grown under As-rich conditions, exhibiting a 2x4 surface reconstruction pattern in RHEED. With the initiation of the AlAs layer deposition, the RHEED pattern transitioned to a 2x2 pattern. The commencement of LT InSb growth saw the complete disappearance of the streaky RHEED pattern from AlAs within approximately 20 seconds, replaced by a pattern of dots indicating surface faceting. While LT InSb continued to grow, the dots gradually faded, and arcs emerged in the pattern, suggesting some amorphous growth. After heating the substrate to a temperature of 350 °C, the RHEED pattern recovered, exhibiting 1x3 reconstruction streaks. Upon the initiation of HT InSb growth, the RHEED pattern transitioned to a 2x2 pattern and persisted in that state until the completion of the process.

\section{Surface Morphology}

\begin{figure*}
     \centering
     \begin{subfigure}{0.6\textwidth}
         \centering
         \includegraphics[width=\textwidth]{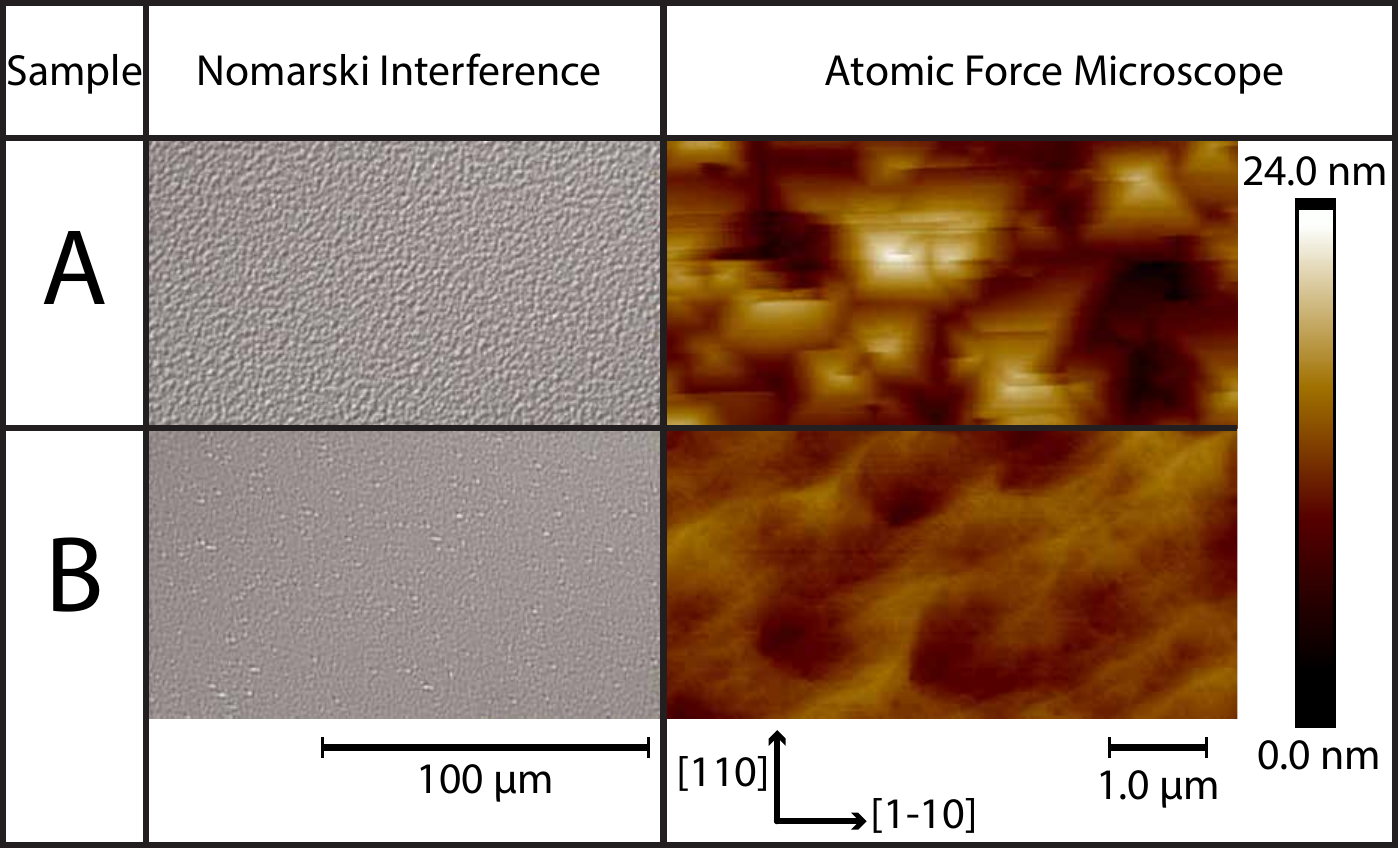}
         \caption{ }
         \label{fig:Nomarski_AFM}
     \end{subfigure}
     \hfill
     \begin{subfigure}{0.39\textwidth}
         \centering
         \includegraphics[width=\textwidth]{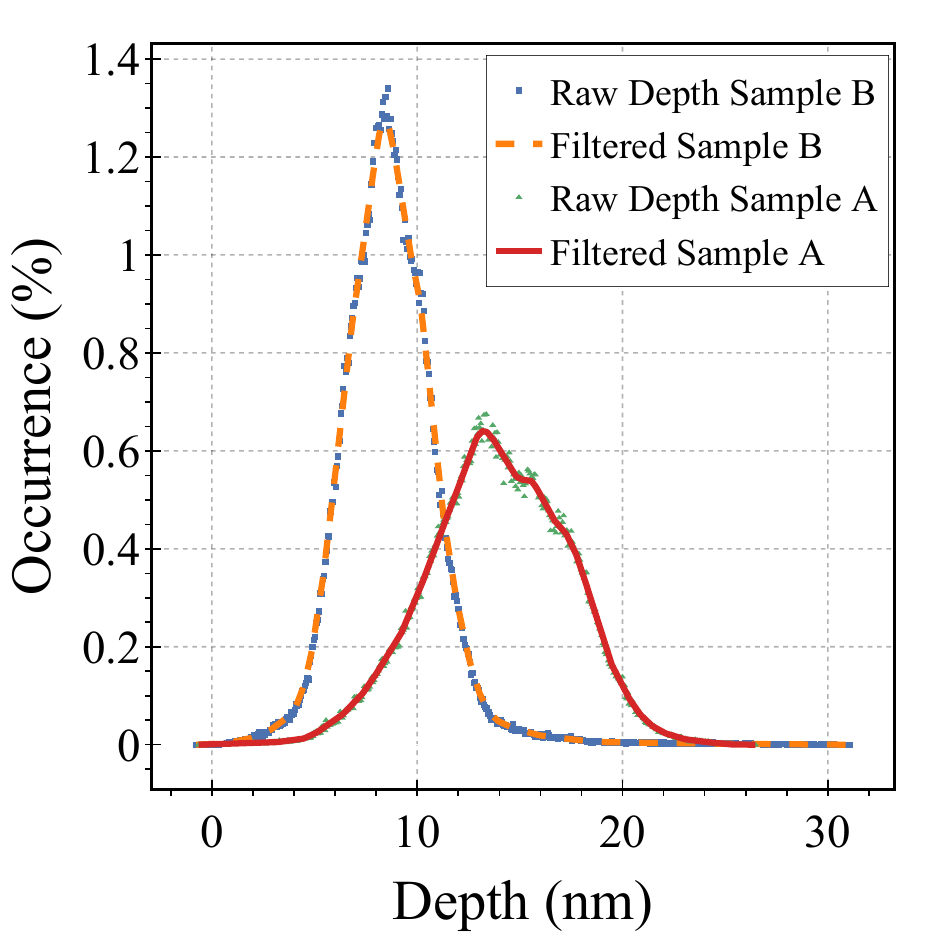}
         \caption{ }
         \label{fig:AFM_Histogram}
     \end{subfigure}
     \caption{(a) A table showing images of NIC images alongside AFM images of both samples A and B. (b) Depth histograms for the AFM images in (a).}
     \label{fig:AFM_Scans}
\end{figure*}

Following the growth process, the wafers underwent examination using a Nomarski interference contrast (NIC) microscope, revealing the presence of 3D features on the surface of both wafers. Then, samples cleaved from these wafers were subjected to further scrutiny using an atomic force microscope (AFM). Figure \ref{fig:Nomarski_AFM} illustrates the surfaces of samples extracted from the center of both wafers. The NIC microscope observations indicate that the surface of Sample B exhibits a smoother texture compared to Sample A. Consistent with this observation, AFM images reveal that the surface of Sample A is covered with pyramidal hillocks, while the surface of Sample B lacks distinguishable hillocks. The formation of hillocks on the surface of Sample B has been effectively suppressed, resulting in a significantly smoother surface with a lower root mean square (RMS) roughness. The AFM scans cover a region of 10.0 µm$^2$, with an RMS roughness of 3.38 nm for Sample A and 2.32 nm for Sample B. The depth histograms in Figure \ref{fig:AFM_Histogram} further emphasize the difference in surface morphology. The lines in Figure \ref{fig:AFM_Histogram} are generated by applying a low-pass Gaussian filter to the raw depth data, effectively smoothing out the data.

\section{\label{sec:methods} Measurements of Crystal Defects}

To qualitatively assess and quantitatively enumerate the types of defects present in the InSb crystals, we employed XRD and electron channeling contrast imaging (ECCI) techniques. XRD measurements were conducted using two distinct diffractometers. The initial diffractometer employed was a Bruker QC3 system equipped with an X-ray tube featuring a copper target and a scintillation detector. This system generated a beam with a divergence of less than 10'' at a wavelength of 1.54 Å, operating in a triple-axis mode. In this mode, a two-bounce channel cut silicon (022) analyzer crystal constrained the detector acceptance angle to 12''. Later on, synchrotron radiation from the BXDS-IVU beamline at the Canadian Light Source (CLS) was employed for the 3D reciprocal space studies. This setup produced a beam with a divergence of 105'' at a wavelength of 1.25 Å, utilizing a detector consisting of a 487 x 195 pixel array with an angular acceptance of 37.8'' for each pixel.

Additionally, a scanning electron microscope (SEM) was utilized in a configuration where backscattered electron signals were measured in ECCI. The SEM operated at an accelerating voltage of 20 kV, and images were captured at a magnification of 20 000 times.

\subsection{\label{sec:X-Ray} X-ray Diffraction}

\subsubsection{\label{sec:Rocking} Rocking Curves}

TDD is measured as the number of threading dislocations that pass through a chosen cross-section plane of the crystal, typically this is the growth plane. TDD is measured in units of cm$^{-2}$. One method for measuring TDD uses XRD rocking curve ($\omega$) scans to measure the width of the Bragg peak associated with a specific set of crystalline planes. Each dislocation induces strain in the crystal structure, leading to localized tilting of the crystal planes in the region surrounding the dislocation. Consequently, there is broadening of the Bragg peak in $\omega$. In a triple-axis configuration, a diffractometer can measure light intensity for these tilts with high angular resolution, allowing determination of the Bragg peak width, from which an estimate for TDD can be derived.\cite{RN254}

Equation (\ref{eq:FWHM}) presents a model for calculating the full width at half-maximum (FWHM) of the Bragg peak, denoted as $\beta_m (hkl)$. Several factors contribute to FWHM broadening, most of which can be eliminated with reasonable assumptions. Even in a dislocation-free crystal, some intrinsic broadening persists. $\beta_0 (hkl)$ and $\beta_d (hkl)$ represent intrinsic broadening from the sample and the diffractometer's conditioning crystals, respectively. Generally, intrinsic broadening in cubic crystals is less than 10'',\cite{RN254} which is below the divergence of the considered diffractometer, thus negligible. $\beta_L (hkl)$ accounts for broadening due to crystal size and is inversely proportional to the thickness of the layer being examined. This factor can be disregarded for layers thicker than 1 $\mu$m.\cite{RN254} Similarly, $\beta_r (hkl)$, associated with curvature of the sample, can be ignored as the InSb layer is fully relaxed, exhibiting minimal residual strain insufficient to induce curvature. The factor $\beta_{\varepsilon}(hkl)$ is linked to strain induced by threading dislocations, and the final factor, $\beta_{\alpha}(hkl)$, results from the tilt caused by threading dislocations.

\begin{eqnarray}
\beta^2_m (hkl) = && \beta^2_0 (hkl) + \beta^2_d (hkl) + \beta^2_{\alpha} (hkl) + \beta^2_{\varepsilon} (hkl)+\nonumber \\
&&  \beta^2_L (hkl) + \beta^2_r (hkl)
\label{eq:FWHM}
\end{eqnarray}

In Ref. \onlinecite{RN254}, the estimation of TDD required three rocking curves from distinct Bragg angles. This was because $\beta_{\varepsilon}(hkl)$ and $\beta_{\alpha}(hkl)$ remain as influencing factors, and a line with at least three points was needed to fit and separate these factors.

With our diffractometer, this method can be modified so that TDD can be estimated with a single rocking curve measurement. The key difference here is that our scans are done with an analyzer crystal which allows the diffractometer to operate in triple-axis mode with reduced acceptance angle.

A detector with a larger acceptance angle captures X-rays corresponding to a range of $2\theta$ values and thus measures parts of the crystal with different strains as if they were all the same strain. Simply put, the detector `window' is larger and so the crystal has to be rotated further along the $\omega$ axis before the Bragg peak is `out of view'. This effect is indistinguishable from the effect of tilted planes which would also make the Bragg peak appear broader while rocking $\omega$. Reducing the detector's acceptance angle diminishes the effects from different strains but doesn't affect the broadening caused by tilted planes. This is why, in triple-axis mode, the FWHM of a Bragg peak in a rocking curve scan is not sensitive to the effects of strain from threading dislocations and $\beta_{\varepsilon}(hkl)$ can be neglected. Therefore, the FWHM of the rocking curve can directly measure the tilt caused by threading dislocations, as expressed by the relationship $\beta^2_{m}=\beta^2_{\alpha}.$

Equation (\ref{eq:FWHM3}) can be used to calculate a TDD estimate directly from the FWHM of a triple-axis scan. $D$ is the TDD and $b$ is the length of the Burgers vector for these dislocations. InSb forms a zincblende crystalline structure with a lattice constant of $a = 6.479$ Å. When a zincblende crystal structure is grown on a (001) surface, all dislocations exhibit Burgers vectors of the form $\mathbf{b} = \frac{1}{2} a \langle 110 \rangle $.\cite{RN273} So, the length of the Burgers vector is always $b=4.581$ Å. 

\begin{equation}
\beta^2_{\alpha} (hkl)  = 2\pi \ln 2 b^2 D
\label{eq:FWHM3}
\end{equation}

An additional characteristic observed in crystals with dislocations is anisotropy in the FWHM of the Bragg peak.  To assess the quality of crystals grown on Wafer A and Wafer B, 12 different rocking curve scans were made at different azimuthal angles, spaced evenly by $30^{\circ}$. A Gaussian function as defined in, Eq. (\ref{eq:Gauss}), was used to fit the intensity of the Bragg peak, $I(\omega)$, of each rocking curve. Each of the calculated fits matched the experimental data with a coefficient of determination, $R^2$, greater than 0.99. Once the $\sigma$ value was determined, the FWHM could be calculated from Eq. (\ref{eq:cFWHM}), where the index $i$ denotes scans from different directions. Subsequently, Eq. (\ref{eq:FWHM3}) was used to calculate the TDD estimate.

\begin{equation}
I(\omega) = A \exp\left(\frac{-(\omega-\omega_0)^2}{2\sigma^2}\right) + C
\label{eq:Gauss}
\end{equation}

\begin{equation}
FWHM_i = 2\sqrt{2\ln(2)}\sigma_i
\label{eq:cFWHM}
\end{equation}

\begin{figure}
\centering
\includegraphics[width=0.48\textwidth]{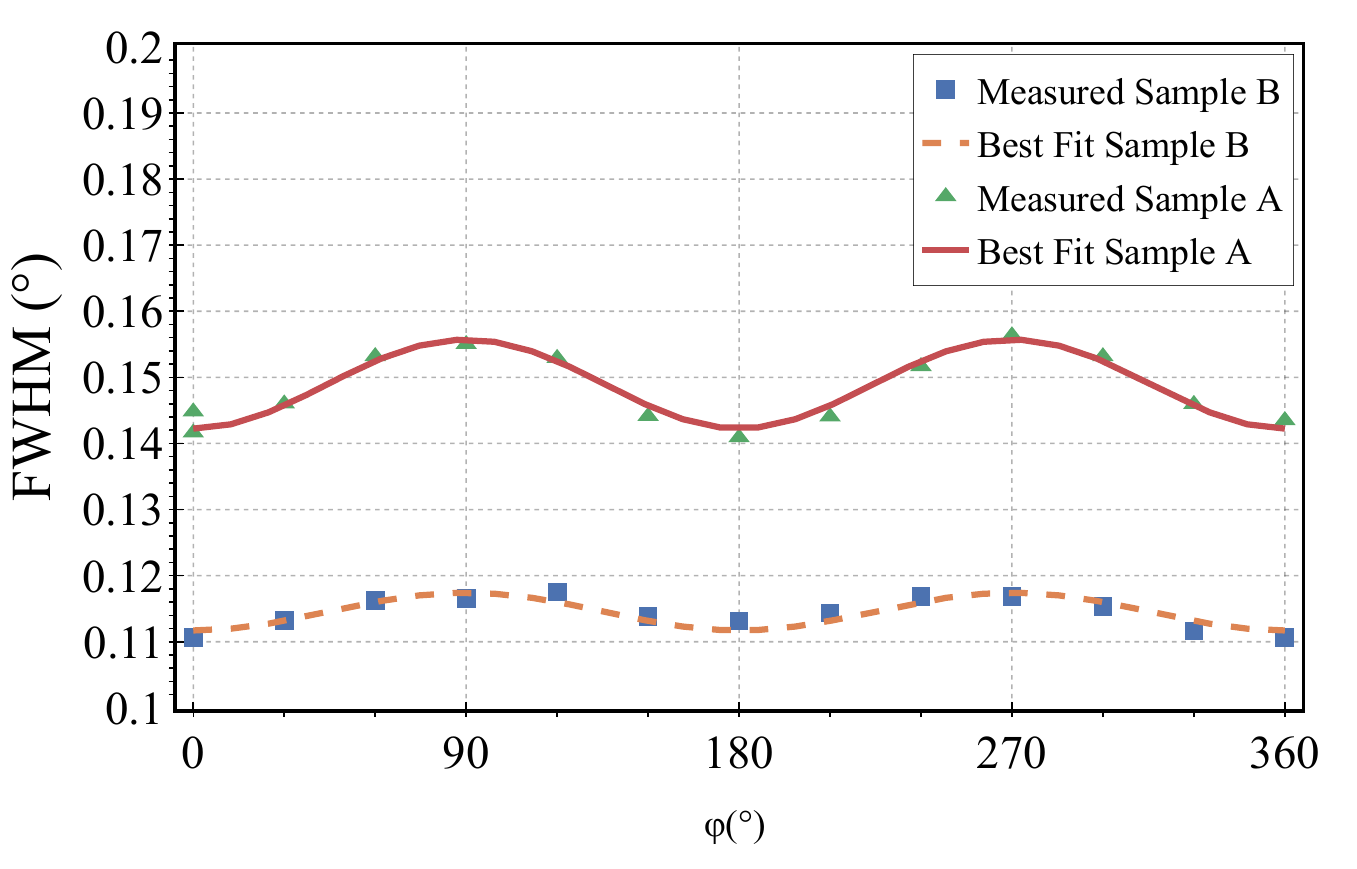}
\caption{The FWHM measured for each rocking curve scan vs. the corresponding azimuthal angle $\varphi$. When $\varphi = 0^{\circ}$, the X-ray beam travels in the [$110$] direction, and when $\varphi = 270^{\circ}$, it travels in the [$1\bar{1}0$] direction.}
\label{fig:TA_RC}
\end{figure}

Figure \ref{fig:TA_RC} shows the calculated FWHM for each scan versus the corresponding azimuthal angle. These FWHM values oscillate with $\varphi$ and the dependence on $\varphi$ was fit by Eq. (\ref{eq:phi}). The fitting parameters in Eq. (\ref{eq:phi}) are $D_{\alpha}$ and $D_{\beta}$, density of $\alpha$ and $\beta$ type dislocations which will be described below. To assess the fit quality, the coefficient of determination, $R^2$, was calculated. The fitting parameters are shown in Table \ref{tab:phi}.

\begin{equation}
\beta^2_{m} (hkl)  = 2\pi \ln 2 b^2 \left[\cos^2(\varphi) D_{\alpha} + \sin^2(\varphi) D_{\beta} \right]
\label{eq:phi}
\end{equation}

\begin{table}
\caption{Results from fitting FWHM data to Eq. (\ref{eq:phi})}
\label{tab:phi}
\begin{ruledtabular}  
\begin{tabular}{cccc}
Parameter						&Units			& Sample A		& Sample B		\\
\hline
$D_{\alpha}$					&$cm^{-2}$		& 6.741E08		& 4.157E08 		\\
$D_{\beta}$						&$cm^{-2}$		& 8.078E08		& 4.592E08		\\
$\frac{D_{\alpha}}{D_{\beta}}$	&$-$			& 0.8345		& 0.9053		\\
$R^2$							&$-$			& 0.9499		& 0.8104		\\
\end{tabular}
\end{ruledtabular}
\end{table}

\begin{figure}
     \centering
      \includegraphics[width=0.48\textwidth]{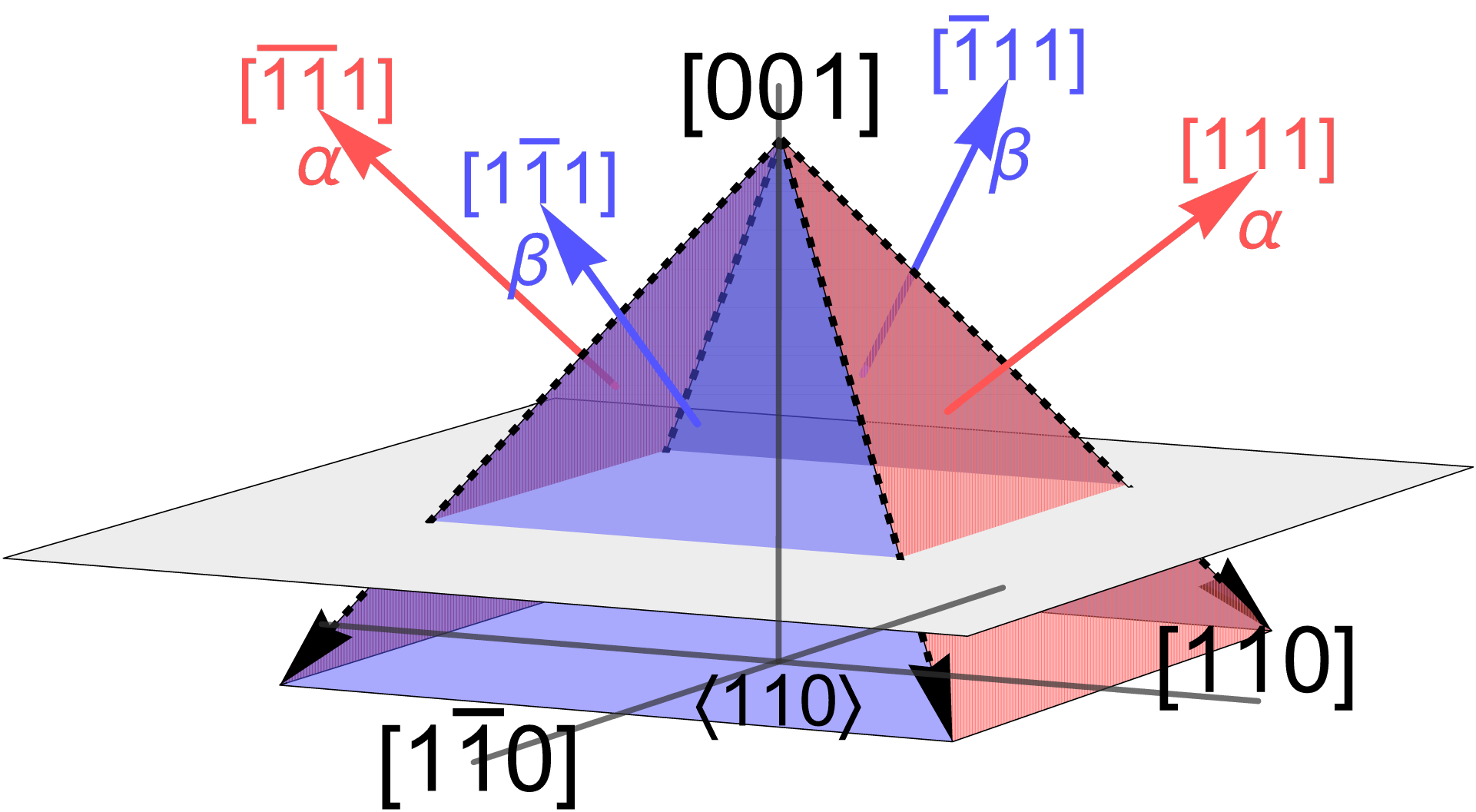}
     \caption{A graphical representation of the active slip systems that intersect the (001) plane in a zincblende crystal structure. The triangular faces of the pyramid represent the \{111\} glide planes and the edges of these faces that intersect the (001) plane are along the $\langle 110 \rangle$ Burgers vectors.}
	  \label{fig:Slips}
\end{figure}

This dependence on $\varphi$ is likely caused by an unequal distribution of dislocations on different slip systems within the crystal structure.\cite{RN255, RN262, RN274} It is important to clarify that measuring the tilt of the (004) planes is equivalent to measuring the tilt of the (001) planes. For a zincblende crystal, such as InSb, all dislocations on the (001) plane form $\frac{1}{2}a \langle 110 \rangle \{ 111 \}$ slip systems,\cite{RN273} where $a$ is the cubic lattice constant. Figure \ref{fig:Slips} shows a graphical representation of these slip systems intersecting the (001) plane. 

In Figure \ref{fig:Slips}, a pyramid intersects the (001) plane. The four dashed edges of the pyramid, each marked with an arrow, represent different $\langle 110 \rangle$ Burgers vectors. Additionally, each of the four triangular faces of the pyramid represents a different $ \{ 111 \}$ glide plane. A slip system is defined by the pairing of a glide plane and a Burgers vector that is contained within that plane. Each of the glide planes depicted in Figure \ref{fig:Slips} contains two Burgers vectors that intersect the (001) plane. So, eight slip systems intersect the (001) plane. 

We can categorize these slip systems as two types based on their glide planes. If the glide planes contain the $[1\bar{1}0]$ direction, then the dislocations that slide on these planes are called $\alpha$-type dislocations. Similarly, when the glide planes contain the $[110]$ direction, the dislocations are called $\beta$-type. The $\alpha$ and $\beta$ type dislocations have line vectors in the $[1\bar{1}0]$ and $[110]$ directions, respectively.

When these types of dislocations exist in unequal proportions, the FWHM of a rocking curve scan will change depending on the azimuthal angle $\varphi$. This is because the dislocations cause tilt about the axis of the line vector.\cite{RN273} A rocking curve scan is sensitive only to the tilt about the rocking axis, $\omega$. When $\varphi=0^{\circ}$, the $\omega$ axis aligns with the $[1\bar{1}0]$ direction, making this scan most sensitive to the tilt induced by $\alpha$-type dislocations. Conversely, at $\varphi=90^{\circ}$, the measured FWHM is primarily influenced by $\beta$-type dislocations. At any other angle, the measured FWHM results from the superposition of the two effects. The fitted values $D_{\alpha}$ and $D_{\beta}$ from Table \ref{tab:phi} correspond to the TDD estimates from the minimum and maximum FWHMs in Figure \ref{fig:TA_RC}, respectively.

\subsubsection{\label{sec:Polar_method} Reciprocal Space Polar Visualization}

In a separate set of experiments, an X-ray array detector and synchrotron radiation were used to perform symmetrical $\omega$-$2\theta$ scans of the (004) Bragg peak for InSb for both samples A and B. The scans were repeated for 12 different azimuthal angles, $\varphi$, spaced evenly by 30$^{\circ}$. An angle of $\varphi = 0^{\circ}$ has the X-ray beam travelling along the $[110]$ direction and $\varphi = 270^{\circ}$ corresponds to $[1\bar{1}0]$. Multiple scans at different azimuths reveal the anisotropic nature of the defects within the crystal. 

\begin{figure}
     \centering
     \begin{subfigure}{0.4\textwidth}
         \centering
         \includegraphics[width=\textwidth]{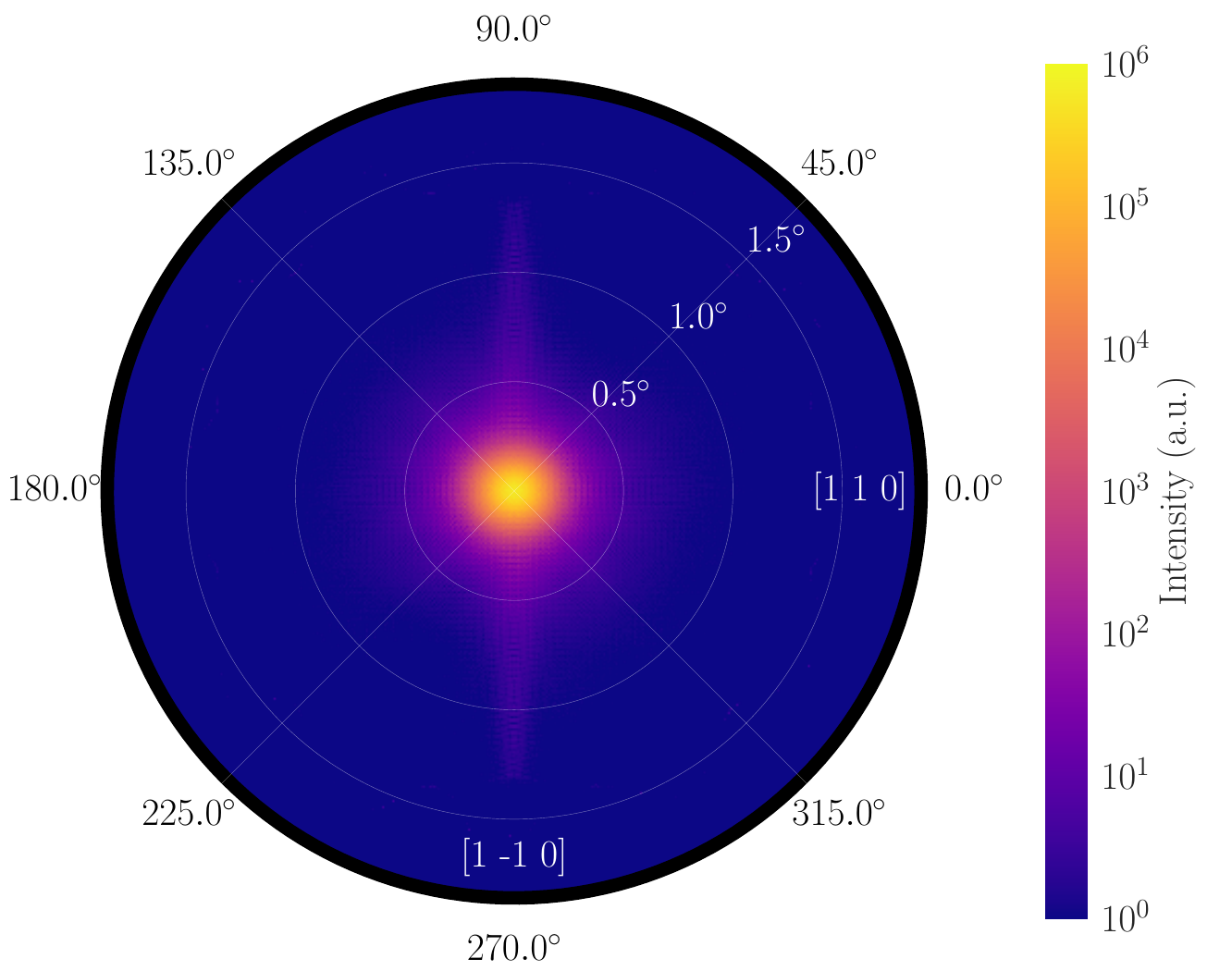}
		 \caption{}
         \label{fig:Pole_Data_A}
     \end{subfigure}
  	\hfill
     \begin{subfigure}{0.4\textwidth}
         \centering
         \includegraphics[width=\textwidth]{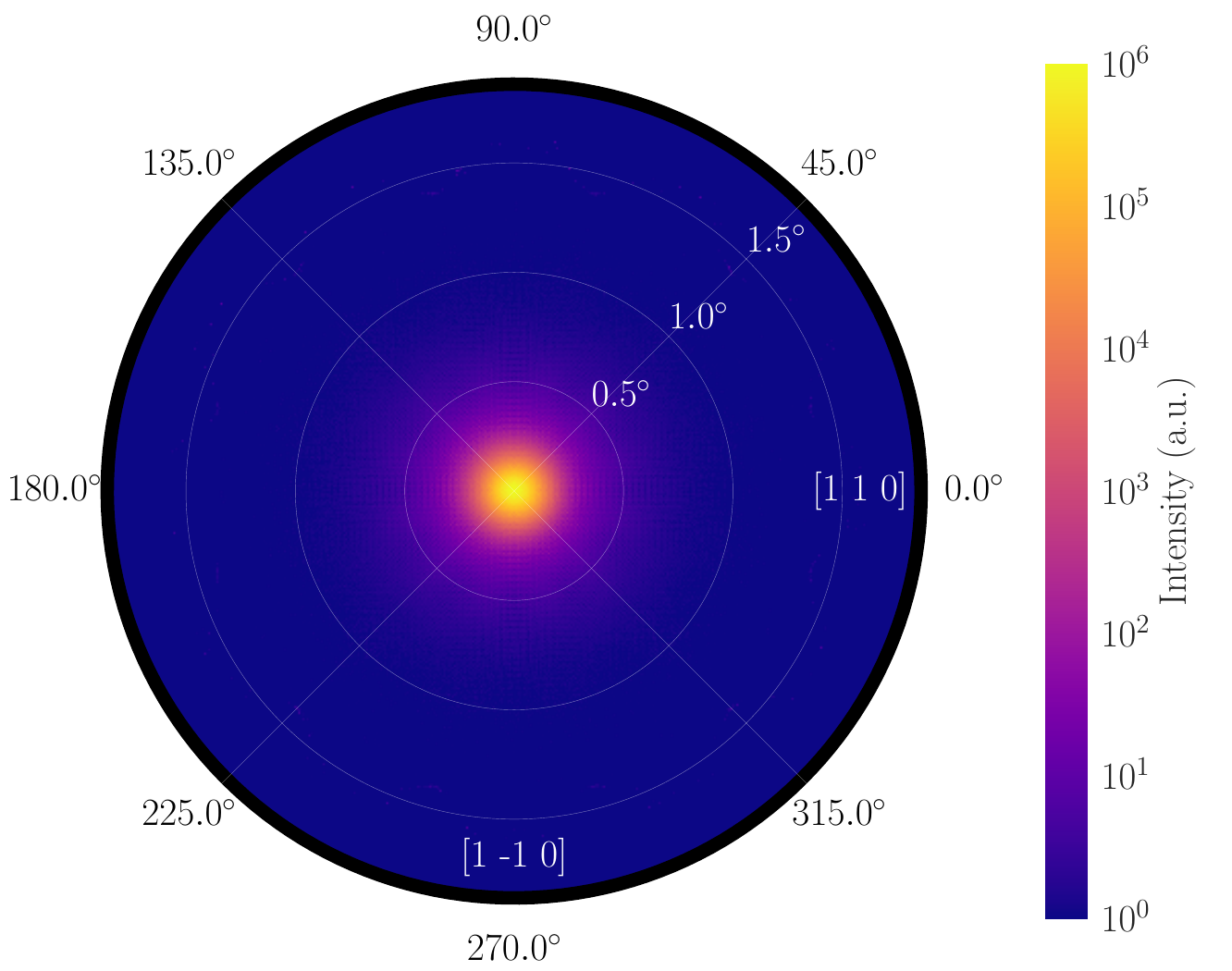}
         \caption{}
         \label{fig:Pole_Data_B}
     \end{subfigure}
     \hfill
     \begin{subfigure}{0.4\textwidth}
         \centering
         \includegraphics[width=\textwidth]{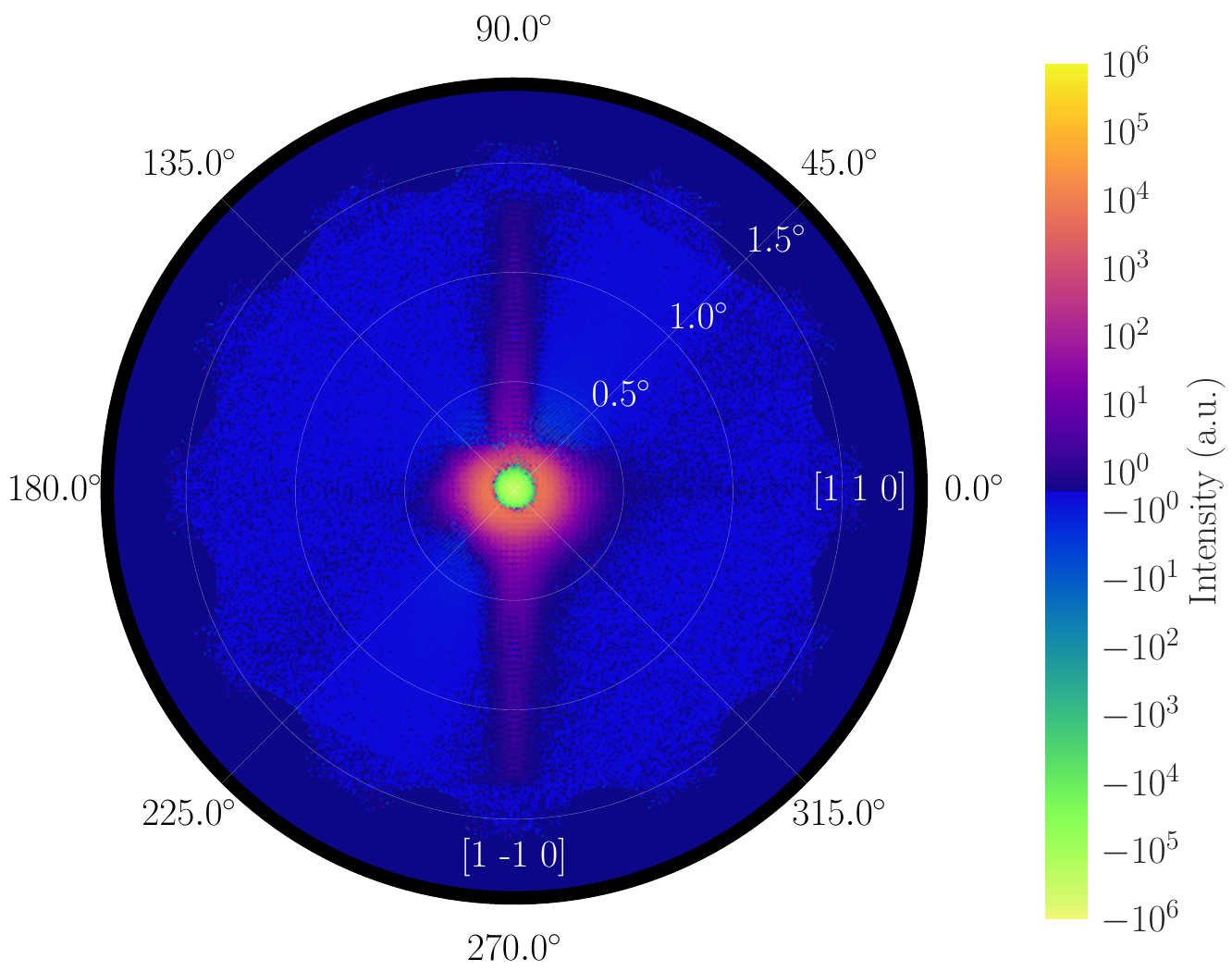}
         \caption{}
         \label{fig:Data_A-B}
     \end{subfigure}
     \caption{(a) Sample A with nominal lattice spacing 1.62 Å +/- 1.70\% (b) Sample B with nominal lattice spacing 1.62 Å +/- 1.70\%. (c) This image is obtained by applying a symmetric logarithmic color scale to the result of subtracting the image in (b) from the image in (a). All pixel intensities are in arbitrary units and are normalized equally between all images.}
	  \label{fig:Data}
\end{figure}

Using appropriate matrix operations, wave vectors of diffracted X-rays, corresponding to each pixel, in each image, during every scan, were calculated and from those, the related scattering vectors, $\mathbf{Q}$, were calculated. Thus, all of the real space positions of each pixel were mapped to coordinate points in reciprocal space.\cite{RN423} Each 3D reciprocal space map (RSM) at different azimuthal angles included some new volume and some volume of space that overlapped with the other RSMs. All of the different azimuthal RSMs were combined to cover a larger region of reciprocal space around the 004 coordinate point. Subsequently, a change of basis from Cartesian coordinates to polar spherical coordinates was applied to each data point.  This transformation allows the generation of graphical representations similar to pole figures, as illustrated in Figure \ref{fig:Data}.

\begin{figure}
     \centering
     \includegraphics[width=0.48\textwidth]{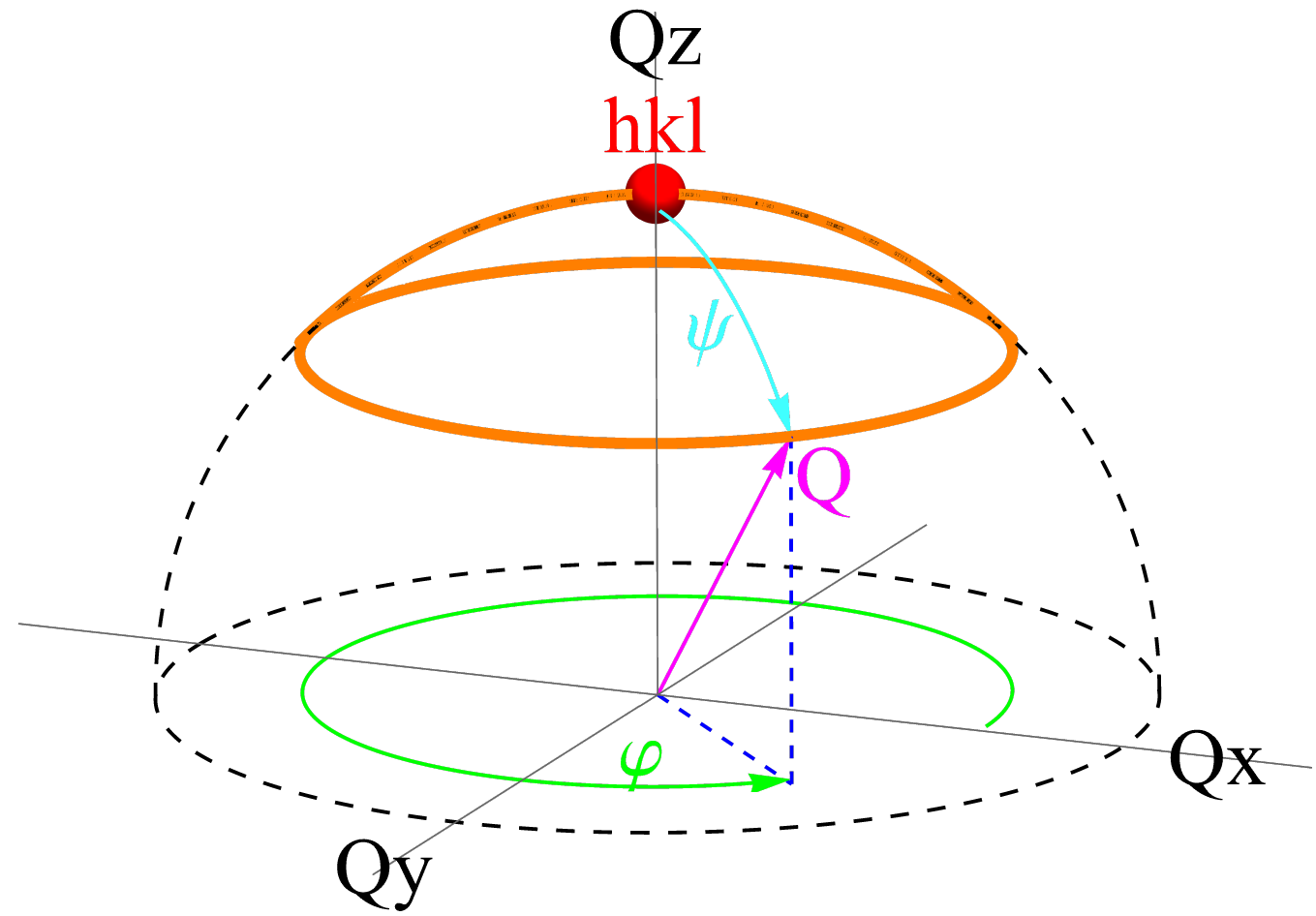}
     \caption{A schematic representation of the region of reciprocal space that is mapped in the polar plots. The reciprocal space point $hkl$ is indicated by the small sphere. $\psi$ is the magnitude of the tilt and $\varphi$ indicates the direction of the tilt for the corresponding scattering vector, $\mathbf{Q}$.}
	  \label{fig:Sphere}
\end{figure}

To understand what region of reciprocal space is shown in these polar plots, consider the scattering vector $\mathbf{Q}$ as shown in Figure \ref{fig:Sphere}. If $\mathbf{Q}$ is freely pivoted around the origin such that $0\le\psi\le90^{\circ}$ and $0\le\varphi<360^{\circ}$ without the vector's length changing, tracing the tip of $\mathbf{Q}$ will draw a hemisphere in reciprocal space. This hemisphere represents all possible scattering vectors for planes with a specific lattice spacing corresponding to the length of $\mathbf{Q}$. If we now restrict $\psi$ to $0\le\psi\le c$ where $c<90^{\circ}$ then the area we trace will be a spherical cap as indicated by the thicker lines in Figure \ref{fig:Sphere}. Projecting this region onto a 2D plane yields a circle. As long as the scattering vector is of the correct length to satisfy the Bragg condition for a set of planes ($hkl$), the $hkl$ Bragg peak will appear at the center of this circle.

In Figure \ref{fig:Data}, $hkl$ corresponds to 004. The X-ray detector measures the light intensity on this spherical cap, which is then plotted on a 2D circle. The coordinates of the points in the measured 3D RSM correspond to a set of scattering vectors with discrete lengths. To create the polar plots, a subset of this data must be chosen such that we are only plotting data which corresponds to scattering vector lengths close to the nominal value for the $hkl$ peak of interest.

To achieve this, we define an interval of lattice spacings, $a_{hkl}'-\Delta\le a \le a_{hkl}'+\Delta$, to determine if data is accepted in the subset. Here, $a$ represents the lattice spacing corresponding to the length of the scattering vector for a data point in the 3D RSM, and $a_{hkl}'$ is the nominal lattice spacing around which the interval is centered. The $'$ denotes that this value is arbitrary and doesn't need to correspond exactly to ($hkl$) planes but potentially corresponds to any lattice spacing. The parameter $\Delta$ determines the maximum allowed variation from $a_{hkl}'$. Although the interval is centered around $a_{hkl}'$, an arbitrary central value above or below $a_{hkl}'$ can be selected instead, thereby choosing data corresponding to slightly smaller or larger lattice spacings. In this context, the presence of any X-ray intensity indicates parts of the crystal experience compressive or tensile strain.

The significance of $a_{hkl}'$ is that its value determines what lattice spacing all of the selected data points correspond to. For InSb, the nominal lattice spacing of the (004) planes would be 1.62 Å. Additionally, two coordinates indicating direction and magnitude of tilt, $\varphi$ and $\psi$ respectively, can be read directly from these plots. $\psi$ is measured as the radius, and $\varphi$ is the angle measured counter-clockwise from the positive x axis. $\psi$ corresponds to the amount of tilt away from the exact (004) planes, while $\varphi$ corresponds to the azimuthal angle, indicating the direction of the tilt.

The intensities of all of the polar plots are normalized to the same scale; the relative intensities of points on the plots reveal what proportion of the crystal structure exists with a specific tilt and specific lattice spacing. This unique approach to representing reciprocal space allows us to visualize the distribution of tilted planes, within a crystal, that have a specific lattice spacing with a corresponding level of strain. Thus, we can study the changes in intensity as we vary strain or examine different tilts. This novel technique will be referred to as Reciprocal Space Polar Visualization (RSPV).

\subsubsection{\label{sec:Polar_analysis} Polar Plot Analysis}

Using RSPV, Figures \ref{fig:Pole_Data_A} and \ref{fig:Pole_Data_B} are generated for the 004 Bragg peak of InSb and correspond to Sample A and Sample B, respectively. For all of the data in these two plots, $a$ is between 1.59 Å and 1.65 Å. This range is centered on the nominal lattice spacing between (004) planes for fully relaxed InSb, with a tolerance of $\Delta = 1.70\%$ of 1.62 Å. The same $\Delta$ value was used for all of the polar plots in this article. This tolerance was chosen because it was the smallest value that could be used before obvious artifacts, due to the sampling of discrete data, began to appear.

In both plots, there is a central bright spot, representing the expected 004 Bragg peak. A subtle yet important difference exists between the plot for Sample A and that for Sample B. Sample A has a streak of higher intensity along a line parallel to the $[1\bar{1}0]$ direction, intersecting the main bright spot at the origin. This line appears only in scans where the X-ray beam is parallel to the $[1\bar{1}0]$ direction. Figure \ref{fig:Data_A-B} emphasizes this feature by calculating the difference in intensity between the two samples (Sample A minus Sample B). The colour bar is extended into negative values using a symmetric logarithmic colour scale, allowing us to observe where the data from Sample A is brighter than Sample B and vice versa. In this case, the vertical streak from Sample A is much more apparent. You will also notice that the very center of the plot is completely dominated by Sample B, whereas Sample A is stronger in the region outside of the center with slightly more tilt.

To parametrize the shape and compare results quantitatively, both datasets from Sample A and Sample B were fit to a 2D Gaussian function, where the intensity was calculated with Eq. (\ref{eq:Linear_Fit}). The results from the fitting process are shown in Table \ref{tab:Fitting}. All the assumptions consistent with the previous diffractometer still hold in this scenario. However, $\beta_{\varepsilon}(hkl)$ is excluded for a distinct reason in this case. This exclusion is warranted because, in these plots, every data point corresponds to the same strain value. So, $D_x$ and $D_y$ were calculated using Eq. (\ref{eq:FWHM3}).

\begin{equation}
I = A \exp\left(\frac{-(x-x_0)^2}{2\sigma_x^2}\right) \exp\left({\frac{-(y-y_0)^2}{2\sigma_y^2}}\right)+C
\label{eq:Linear_Fit}
\end{equation}

\begin{table}
\caption{Results from fitting the polar plot data to Eq. (\ref{eq:Linear_Fit}).}
\label{tab:Fitting}
\begin{ruledtabular}
\begin{tabular}{cccc}
Parameter				&Units			& Sample A		& Sample B		\\
\hline
$A$						&a.u.			& 6.22E-01		& 9.69E-01		\\
$x_0$					&${\circ}$		& -4.02E-03		& -8.00E-04		\\
$y_0$					&${\circ}$		& 9.48E-04		& 4.15E-04		\\
$\sigma_x$				&${\circ}$		& 5.66E-02		& 4.91E-02		\\
$\sigma_y$				&${\circ}$		& 6.15E-02		& 5.17E-02		\\
$C$						&a.u.			& 1.35E-04		& 1.58E-04		\\
$R^2$					&$-$			& 0.994			& 0.994			\\
FWHM$_x$				&${\circ}$		& 0.133			& 0.116			\\
FWHM$_y$				&${\circ}$		& 0.145			& 0.122			\\
$\epsilon$  			&$-$			& 0.391			& 0.310			\\
$D_x$					&$cm^{-2}$		& 5.93E08		& 4.46E08		\\
$D_y$					&$cm^{-2}$		& 6.99E08		& 4.94E08		\\
\end{tabular}
\end{ruledtabular}
\end{table}

These fitting parameters were calculated for the data on a linear scale. The data has been fit very well and the coefficient of determination values, $R^2$, are greater than $0.99$. However, the fit works well only for the central bright spot and does not include the vertical streak seen from Sample A. Sample B has a maximum intensity, $A$, that is 56\% larger than for Sample A. Using Eq. (\ref{eq:cFWHM}) with $\sigma_x$ and $\sigma_y$, the FWHM has been calculated for both the $x$ and $y$ directions. The FWHM values in Table \ref{tab:Fitting} indicate that the central bright spot for Sample B is smaller and more focussed, hence the large increase in $A$ for Sample B. Also, the FWHM values for Sample B show that shape of the central bright spot has 21\% lower eccentricity, calculated as $\epsilon$ in Eq. (\ref{eq:ecc}). This shows the Bragg peak for Sample B is more isotropic.

\begin{equation}
\epsilon = \sqrt{1-\frac{{FWHM}_y^2}{{FWHM}_x^2}}
\label{eq:ecc}
\end{equation}

\subsection{\label{sec:SEM}Scanning Electron Microscope}

\begin{figure*}
     \centering
     \begin{subfigure}{0.4\textwidth}
         \centering
         \includegraphics[width=\textwidth]{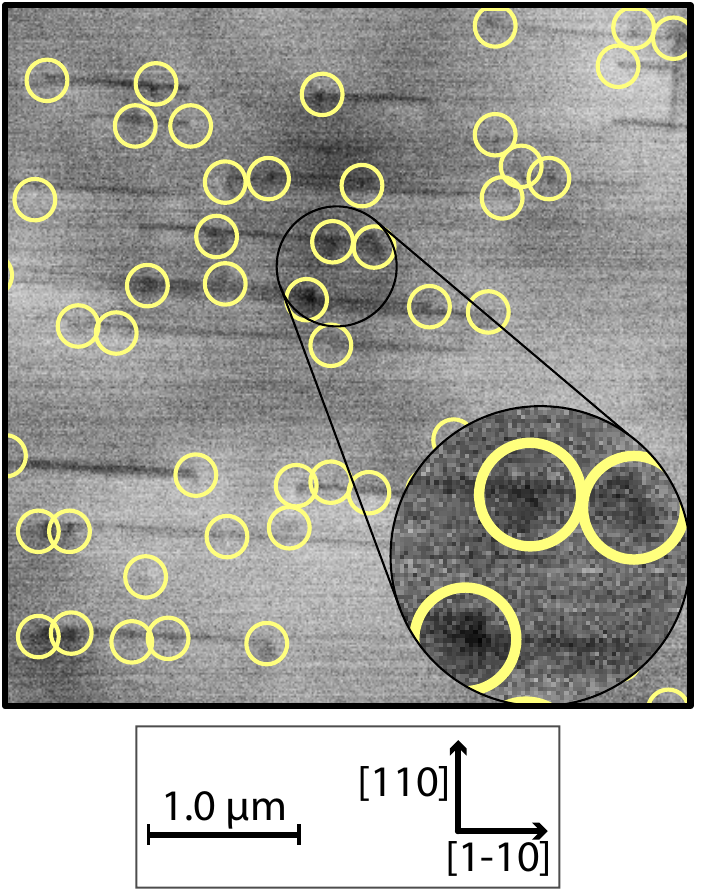}
		 \caption{}
         \label{fig:SEM_G0845}
     \end{subfigure}
     \begin{subfigure}{0.4\textwidth}
         \centering
         \includegraphics[width=\textwidth]{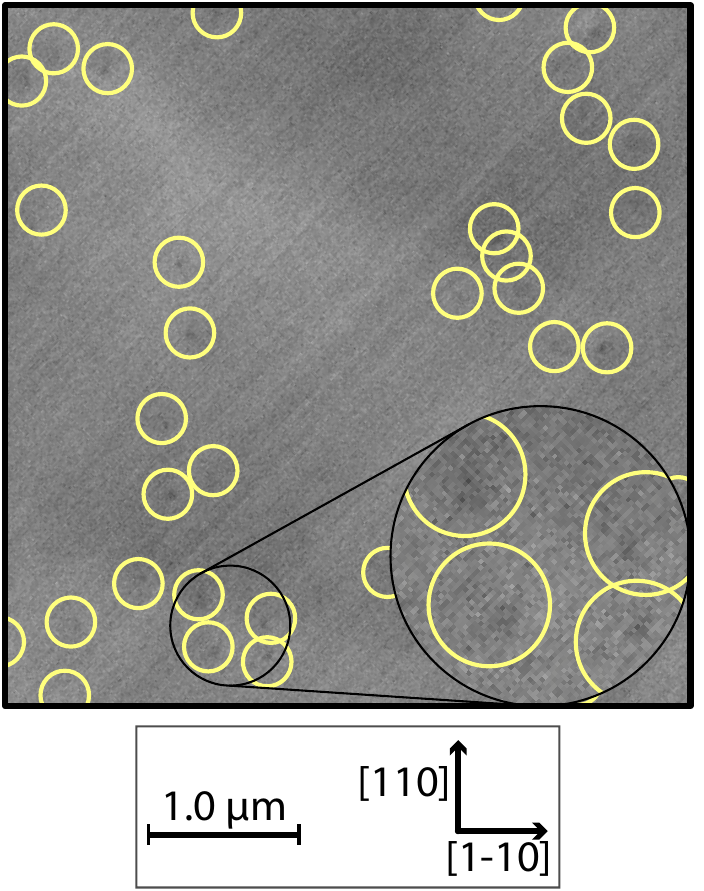}
         \caption{}
         \label{fig:SEM_G0846}
     \end{subfigure}
     \caption{Characteristic sections of SEM ECCI images for (a) sample A and (b) sample B. Small dots of increased contrast caused by TDD are circled. The crystallographic directions and image scale are indicated in the images. The image insets show zoomed regions where threading dislocations were identified.} 
	  \label{fig:ECCI}
\end{figure*}

ECCI was used as another method to estimate the TDD in both samples. This gives a second, independent, result for comparing with the XRD results. ECCI produces a contrast image by measuring the signal from backscattered electrons acquired with an SEM. Figure \ref{fig:ECCI} shows the ECCI results for both samples. Dislocations within a crystal structure induce strain, causing slight changes to the orientation of crystallographic planes in a localized region. Any region that has a different orientation will produce a backscattered electron signal different from the background signal in the on-orientation regions. This is how the contrast image is generated. 

The ECCI scans from both samples show a high density of dots that appear distinctly from the line defects in Figure \ref{fig:SEM_G0845}. The dots appear due to the tilt and strain caused by individual threading dislocations. Thus, by counting the number of dots and normalizing by the surface area, another estimate for the TDD can be made, independent of XRD measurements. 

The dots were identified by visual inspection and are circled in Figure \ref{fig:ECCI}. Due to the low contrast, these TDD estimates come with a high uncertainty, but they should serve well to verify the XRD estimates. When it was challenging to determine whether a small contrast was caused by noise or a dislocation, it was generally marked as a dislocation. This strategy will tend to overcount the number of dislocations rather than under count them. This means that these TDD estimates can be considered as upper limits.

For Sample A, 145 dots were found in an area of 2.209E-07 cm$^{2}$. This gives a TDD estimate of 6.565E08 cm$^{-2}$ for Sample A. In an area of the same size, for Sample B, only 107 dots were found which gave a TDD estimate of 4.844E08 cm$^{-2}$. The imaged areas of the samples are too small to draw conclusions about the uniformity of TDD across the sample surfaces, and these results might not accurately represent the average TDD. Furthermore, the TDD may not be uniform along the growth direction, and ECCI is only sensitive to TDD near the surface because the electron beam will not penetrate the entire depth of the InSb layer.

\section{\label{sec:Analysis}Analysis of Experimental Results}

\subsection{\label{subsec:TDD}Threading Dislocation Density}

Through three different methods, the TDD of InSb grown on GaAs (001) has been estimated. FWHM measurements of the 004 Bragg peak for InSb suggests that there is an unequal distribution of $\alpha$ and $\beta$ type dislocations that propagate along two sets of glide planes perpendicular to each other. TDD estimates from the XRD rocking curves show that $\beta$-type dislocations are always present in greater proportion than the $\alpha$-type in both samples. This causes the Bragg peak to be broadened more in the $[1\bar{1}0]$ direction than it is in the $[110]$ direction, consistent with the results of the 2D Gaussian fitting to the Bragg peak in the polar plots of Figure \ref{fig:Data}. The TDD was estimated independently from XRD by using ECCI to count the individual dislocations. All of the TDD estimates were in good agreement with each other.

ECCI provided a straightforward way to visualize and explicitly count individual threading dislocations, while XRD methods were theoretical and dependent on assumptions. Combining both methods is advantageous because ECCI can only measure TDD for a small area of the sample and has a high uncertainty, whereas XRD methods are precise and cover a larger part of the crystal but involve complex analysis. The agreement between the two methods ensures the reliability of the results. The largest difference between ECCI and XRD estimates was 18.73\%, observed when comparing $D_{\beta}$ to the ECCI TDD for Sample A. This discrepancy may be attributed to the fact that ECCI only measures TDD in a localized area near the surface of the crystal, while X-rays in XRD illuminate a larger surface area and penetrate the entire depth of the InSb layer, providing more representative results for the entire crystal. Furthermore, the poor contrast in ECCI scans makes identifying threading dislocations difficult, further increasing the uncertainty in ECCI results.

Regardless, all methods for estimating TDD show the same trend. Compared to Sample A, the TDD in Sample B is significantly lower. From the rocking curve measurements, this means a reduction of  38.33\% in $D_{\alpha}$ and a 43.15\% reduction in $D_{\beta}$. Moreover, $\beta$ dislocation densities were reduced more than the $\alpha$ densities, making the Bragg peak for Sample B more isotropic with a 21\% lower eccentricity.

One factor contributing to such unequal distributions of $\alpha$ and $\beta$ type dislocations is $\alpha$-type dislocations are terminated with In atoms, while $\beta$-type dislocations are terminated with Sb atoms. The two types of dislocations can have different nucleation energies and glide velocities, resulting in unequal TDDs.\cite{RN274}

In epitaxial growth, pyramidal hillocks are formed by spiral growth around threading dislocations, creating a series of stepped terraces rising to the tip of the pyramid. The incline of the pyramid terraces has a natural facet angle. A misoriented (a.k.a. offcut or vicinial) substrate surface will also have terraces because the surface is not parallel to the crystallographic planes. If the angle of misorientation matches or exceeds the natural facet angle of the hillocks, then new terraces cannot be grown with the typical spiral mechanism, suppressing the formation of hillocks.\cite{RN168}

The terraced substrate surface might also be promoting misfit dislocation generation at the InSb/Substrate interface, leading to an interfacial misfit dislocation array. This increased presence of misfit dislocations would improve strain relaxation at the interface, reducing TDD generation in the rest of the InSb layer. The reduced threading dislocation density (TDD) in Sample B is thus attributed to the misoriented substrate surface.

\subsection{\label{subsec:Microtwins}Microtwin Defects}

ECCI measurements for Sample A (Figure \ref{fig:SEM_G0845}) also revealed a high density of line defects. Similar line defects were observed in other works where InSb was directly grown on GaAs (001) substrates.\cite{RN168, RN419, RN420, RN421} This kind of defect is known as an oriented abrupt step (OAS), resulting from microtwin defects forming in the crystal. OASs form when microtwin defects in the \{111\} plane terminate at the (001) surface of the crystal.\cite{RN419} These OASs are also visible in the AFM image (Figure \ref{fig:Nomarski_AFM}). Most of these OASs run parallel to the $[1\bar{1}0]$ direction, with only a few observed along $[110]$. Additionally, these OASs are exclusively detected in Sample A, indicating that twinning has been suppressed in Sample B, consistent with previous observations.\cite{RN168}

Another unique feature in Sample A was revealed by RSPV. Figure \ref{fig:Pole_Data_A} displays a feature outside the central bright spot, which is broad along the $[1\bar{1}0]$ direction but less so along the $[110]$ direction. This feature crosses the InSb 004 Bragg peak. We hypothesize that the OASs are responsible for this feature, as it appears along the same direction as the OAS defects. The additional scattering effect is only evident when the plane of incidence is parallel to the OAS defects. Each OAS creates a strained region extending along the entire length of the twin boundary, inducing some amount of tilt in the region. Photons traveling along the length of the boundary have a higher probability of being scattered by the tilted planes compared to photons crossing the line in other directions. Notably, this feature was not discernible from rocking curve measurements alone; it was only discovered through the transformation and plotting of 3D reciprocal space maps as polar plots by using the RSPV technique. This underscores the value of this novel technique.

\subsection{\label{subsec:ST}Strain \& Tilt}

\begin{figure*}
     \centering
     \includegraphics[width=0.92\textwidth]{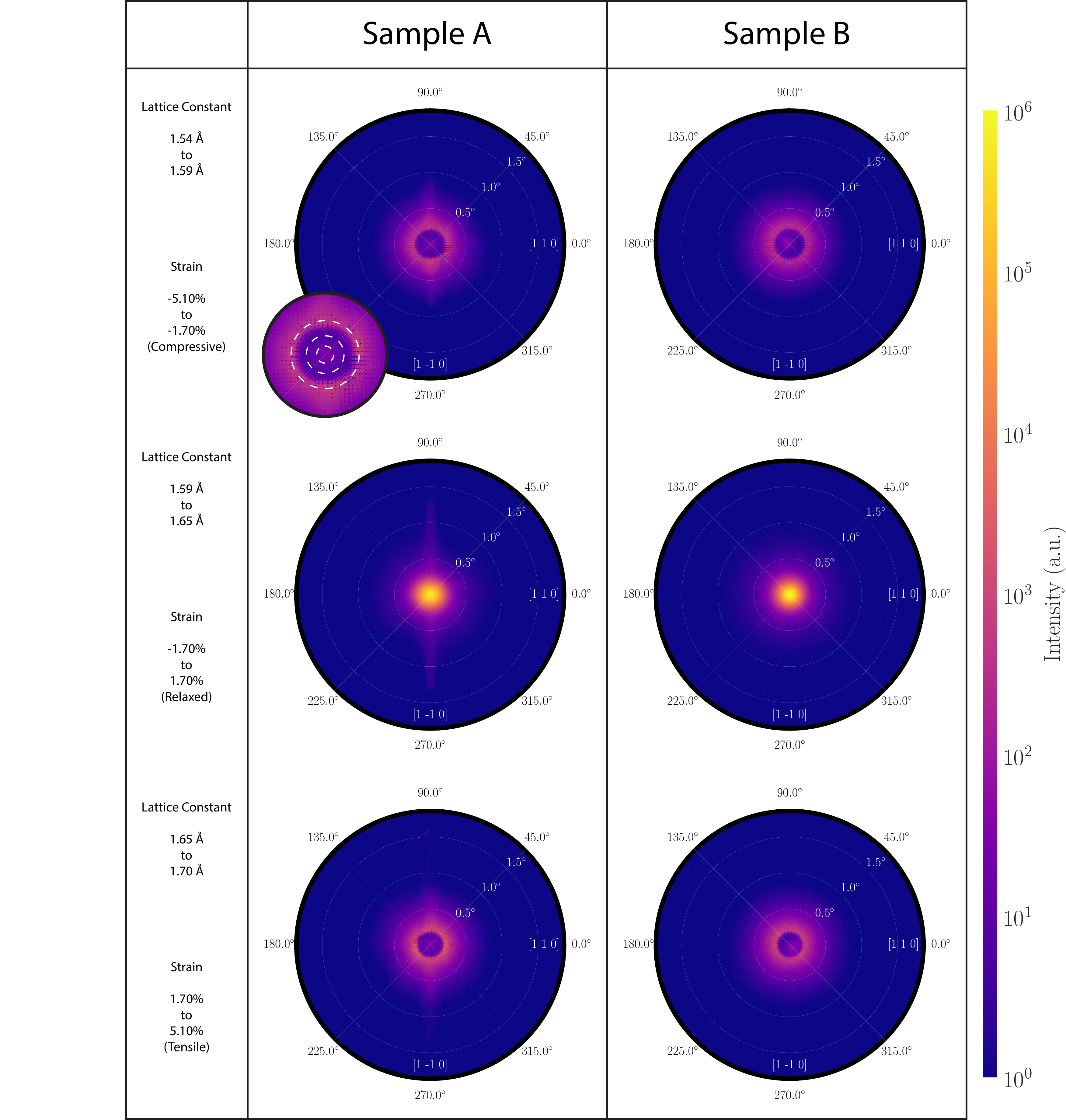}
     \caption{Polar plots of reciprocal space for samples A and B at three different levels of strain. All pixel intensities are in arbitrary units and are normalized equally between all images. The inset on the first row marks 3 regions of intensity within the central area.}
	  \label{fig:Pole_Data}
\end{figure*}

RSPV was used again to create the table of polar plots in Figure \ref{fig:Pole_Data}. Each row of the table corresponds to a different interval of lattice spacings. These values were chosen to ensure no overlap, and each plot represents an entirely exclusive dataset. The polar plots in the first row correspond to lattice spacings $1.54$ Å $\le a < 1.59$ Å. These lattice spacings are smaller than the nominal value for (004) InSb planes, indicating examination of parts of the crystal under compressive strain. The strain ranges from -5.10\% to -1.70\%. The second row of plots in Figure \ref{fig:Pole_Data} correspond to $1.59$ Å $\le a < 1.65$ Å and are identical to the plots given earlier in Figure \ref{fig:Data}. They correspond to the part of the crystal that is within $\pm 1.70\%$ strain which is nearly relaxed. The third row of plots corresponds to $1.65$ Å $\le a < 1.70$ Å with tensile strains ranging from 1.70\% to 5.10\%. 

The reported lattice spacing and strain values exhibit a discrepancy due to rounding decimal places to significant figures. Although the calculation intervals were initially equal, rounding the values for significant figures introduces the perception of a broader range in the relaxed case. The percentage strain values were computed with full precision before rounding for significant figures.

For both Sample A and B, when examining the plots corresponding to compressive or tensile strain, the average intensities in the polar plots are 2 or 5 orders of magnitude lower than that for the relaxed case, respectively. In the plots corresponding to compressive or tensile strain, the central area appears to have three distinct regions of intensity, marked by dashed lines in the inset for the first plot. The very center is significantly dimmer than the outer region and then, moving radially outward, there is an even dimmer ring just outside of the center, followed by a bright ring region. Each plot represents a range of lattice constants yet there is an abrupt boundary between the bright ring and inner dimmer region. This indicates that there is some minimum threshold value for the magnitude of the tilt needed for strained parts of the crystal to form.

For sample A, the streak along $[1\bar{1}0]$ still appears, although it is now limited to about half of its original magnitude of tilt. For sample B, there is a subtle asymmetry in the very center of the compressive and tensile plots. In the compressive plot, the peak intensity at the center of the figure appears slightly tilted towards the $[\bar{1}00]$ direction, while in the tensile plot, the peak intensity is tilted towards the opposite direction, $[100]$. 

The RSPV technique has revealed many previously unobserved features of the reciprocal space of the crystal, and further studies will be needed to identify the cause of these features.

\section{\label{sec:Conclusion}Conclusions}

As revealed by AFM, Sample B exhibited much lower depth variation, while Sample A's surface was covered in pyramidal hillocks. This suppression of hillock formation is attributed to the initial misorientation of the growth surface.

The TDDs for both samples were estimated using two different XRD methods to measure the width of the 004 Bragg peak, and also by counting visible defects in ECCI scans. All of the estimates were in good agreement. The anisotropy of the FWHM of the Bragg peaks in both samples suggests that there are unequal amounts of $\alpha$ and $\beta$ type dislocations. $\beta$ dislocations always had a higher density than $\alpha$ dislocations. It was found that the misoriented sample had a reduction of 38.33\% in $D_{\alpha}$ and a 43.15\% reduction in $D_{\beta}$ compared to Sample A. The reduced TDD suggests that an interfacial misfit dislocation array has more effectively relaxed strain at the InSb/substrate interface in Sample B.

Additionally, the microtwin defects discovered in the ECCI scan for Sample A were found to leave a signature in the RSM causing a streak in the 004 Bragg peak to appear in the RSPV. Neither the microtwin defects nor their XRD signature were present in Sample B. Thus, if we assert that higher quality crystals have lower TDD and more isotropic Bragg peaks, then Sample B was certainly of higher quality than Sample A.

Using the novel method for exploring reciprocal space called, Reciprocal Space Polar Visualization (RSPV), outlined in Section \ref{sec:Polar_method}, it was shown that there is a minimum amount of tilt required for strained parts of the crystal to form. RSPV revealed many subtle features about the shape and intensity of the Bragg peak and how it relates to the various tilts and strains within the crystal. Identifying the cause of some of these features will require further study. Importantly, these features were not distinguishable when using common RSM methods. RSPV provides a way to study a crystal with independent variation of strain or tilt. Furthermore, this method is applicable to any crystal sample or X-ray diffractometer, so long as 3D data can be collected, which will require a 2D X-ray array detector to complete the scans within a reasonable amount of time. Scans with higher spatial resolution may provide a clearer view into the nature of the crystalline defects as well because the width of the intervals for lattice spacings could be reduced further. There are great potential uses for this novel method in future studies, especially when paired with other techniques such as cross-sectional transmission electron microscopy for measuring strain distribution and lattice constants around dislocations.

\begin{acknowledgments}
Part of the research described in this paper was performed at the Canadian Light Source, a national research facility of the University of Saskatchewan, which is supported by the Canada Foundation for Innovation (CFI), the Natural Sciences and Engineering Research Council (NSERC), the National Research Council (NRC), the Canadian Institutes of Health Research (CIHR), the Government of Saskatchewan, and the University of Saskatchewan.

A. Rahemtulla and N. Appathurai, part of the BXDS-IVU beamline team at the CLS, provided invaluable training and support for beamline operation, calibration, and troubleshooting, ensuring smooth data collection.

The University of Waterloo's QNFCF facility was used for this work. This infrastructure would not be possible without the significant contributions of CFREF-TQT, CFI, ISED, the Ontario Ministry of Research \& Innovation and Mike \& Ophelia Lazaridis. Their support is gratefully acknowledged.

In addition to the above institutions, this work was directly supported by the Natural Sciences and Engineering Research Council of Canada (NSERC).
\end{acknowledgments}

\section*{\label{sec:DataAvailability}Data Availability}
The data that support the findings of this study are available from the corresponding author upon reasonable request.


\section*{\label{sec:Declaration}Author Declarations}
The authors have no conflicts to disclose.

\bibliography{Ref}

\begin{thebibliography}{16}%
\makeatletter
\providecommand \@ifxundefined [1]{%
 \@ifx{#1\undefined}
}%
\providecommand \@ifnum [1]{%
 \ifnum #1\expandafter \@firstoftwo
 \else \expandafter \@secondoftwo
 \fi
}%
\providecommand \@ifx [1]{%
 \ifx #1\expandafter \@firstoftwo
 \else \expandafter \@secondoftwo
 \fi
}%
\providecommand \natexlab [1]{#1}%
\providecommand \enquote  [1]{``#1''}%
\providecommand \bibnamefont  [1]{#1}%
\providecommand \bibfnamefont [1]{#1}%
\providecommand \citenamefont [1]{#1}%
\providecommand \href@noop [0]{\@secondoftwo}%
\providecommand \href [0]{\begingroup \@sanitize@url \@href}%
\providecommand \@href[1]{\@@startlink{#1}\@@href}%
\providecommand \@@href[1]{\endgroup#1\@@endlink}%
\providecommand \@sanitize@url [0]{\catcode `\\12\catcode `\$12\catcode `\&12\catcode `\#12\catcode `\^12\catcode `\_12\catcode `\%12\relax}%
\providecommand \@@startlink[1]{}%
\providecommand \@@endlink[0]{}%
\providecommand \url  [0]{\begingroup\@sanitize@url \@url }%
\providecommand \@url [1]{\endgroup\@href {#1}{\urlprefix }}%
\providecommand \urlprefix  [0]{URL }%
\providecommand \Eprint [0]{\href }%
\providecommand \doibase [0]{http://dx.doi.org/}%
\providecommand \selectlanguage [0]{\@gobble}%
\providecommand \bibinfo  [0]{\@secondoftwo}%
\providecommand \bibfield  [0]{\@secondoftwo}%
\providecommand \translation [1]{[#1]}%
\providecommand \BibitemOpen [0]{}%
\providecommand \bibitemStop [0]{}%
\providecommand \bibitemNoStop [0]{.\EOS\space}%
\providecommand \EOS [0]{\spacefactor3000\relax}%
\providecommand \BibitemShut  [1]{\csname bibitem#1\endcsname}%
\let\auto@bib@innerbib\@empty
\bibitem [{\citenamefont {Li}\ \emph {et~al.}(2008)\citenamefont {Li}, \citenamefont {Liu}, \citenamefont {Li}, \citenamefont {You}, \citenamefont {Li}, \citenamefont {Xiong}, \citenamefont {Wang}, \citenamefont {Zhang},\ and\ \citenamefont {Wang}}]{RN412}%
  \BibitemOpen
  \bibfield  {author} {\bibinfo {author} {\bibfnamefont {Z.}~\bibnamefont {Li}}, \bibinfo {author} {\bibfnamefont {G.}~\bibnamefont {Liu}}, \bibinfo {author} {\bibfnamefont {M.}~\bibnamefont {Li}}, \bibinfo {author} {\bibfnamefont {M.}~\bibnamefont {You}}, \bibinfo {author} {\bibfnamefont {L.}~\bibnamefont {Li}}, \bibinfo {author} {\bibfnamefont {M.}~\bibnamefont {Xiong}}, \bibinfo {author} {\bibfnamefont {Y.}~\bibnamefont {Wang}}, \bibinfo {author} {\bibfnamefont {B.}~\bibnamefont {Zhang}}, \ and\ \bibinfo {author} {\bibfnamefont {X.}~\bibnamefont {Wang}},\ }\bibfield  {title} {\enquote {\bibinfo {title} {Thin insb films on gaas substrates by molecular beam epitaxy},}\ }\href {\doibase 10.1143/jjap.47.558} {\bibfield  {journal} {\bibinfo  {journal} {Japanese Journal of Applied Physics}\ }\textbf {\bibinfo {volume} {47}},\ \bibinfo {pages} {558--560} (\bibinfo {year} {2008})}\BibitemShut {NoStop}%
\bibitem [{\citenamefont {D'Costa}\ \emph {et~al.}(2015)\citenamefont {D'Costa}, \citenamefont {Tan}, \citenamefont {Jia}, \citenamefont {Yoon},\ and\ \citenamefont {Yeo}}]{RN413}%
  \BibitemOpen
  \bibfield  {author} {\bibinfo {author} {\bibfnamefont {V.~R.}\ \bibnamefont {D'Costa}}, \bibinfo {author} {\bibfnamefont {K.~H.}\ \bibnamefont {Tan}}, \bibinfo {author} {\bibfnamefont {B.~W.}\ \bibnamefont {Jia}}, \bibinfo {author} {\bibfnamefont {S.~F.}\ \bibnamefont {Yoon}}, \ and\ \bibinfo {author} {\bibfnamefont {Y.-C.}\ \bibnamefont {Yeo}},\ }\bibfield  {title} {\enquote {\bibinfo {title} {Mid-infrared to ultraviolet optical properties of insb grown on gaas by molecular beam epitaxy},}\ }\href {\doibase 10.1063/1.4922586} {\bibfield  {journal} {\bibinfo  {journal} {Journal of Applied Physics}\ }\textbf {\bibinfo {volume} {117}},\ \bibinfo {pages} {223106} (\bibinfo {year} {2015})}\BibitemShut {NoStop}%
\bibitem [{\citenamefont {Zhao}\ \emph {et~al.}(2017)\citenamefont {Zhao}, \citenamefont {Zhang}, \citenamefont {Cui}, \citenamefont {Guan}, \citenamefont {Wang}, \citenamefont {Zhu},\ and\ \citenamefont {Zeng}}]{RN414}%
  \BibitemOpen
  \bibfield  {author} {\bibinfo {author} {\bibfnamefont {X.-M.}\ \bibnamefont {Zhao}}, \bibinfo {author} {\bibfnamefont {Y.}~\bibnamefont {Zhang}}, \bibinfo {author} {\bibfnamefont {L.-J.}\ \bibnamefont {Cui}}, \bibinfo {author} {\bibfnamefont {M.}~\bibnamefont {Guan}}, \bibinfo {author} {\bibfnamefont {B.-Q.}\ \bibnamefont {Wang}}, \bibinfo {author} {\bibfnamefont {Z.-P.}\ \bibnamefont {Zhu}}, \ and\ \bibinfo {author} {\bibfnamefont {Y.-P.}\ \bibnamefont {Zeng}},\ }\bibfield  {title} {\enquote {\bibinfo {title} {Growth and characterization of insb thin films on gaas (001) without any buffer layers by mbe},}\ }\href {\doibase 10.1088/0256-307x/34/7/076105} {\bibfield  {journal} {\bibinfo  {journal} {Chinese Physics Letters}\ }\textbf {\bibinfo {volume} {34}} (\bibinfo {year} {2017}),\ 10.1088/0256-307x/34/7/076105}\BibitemShut {NoStop}%
\bibitem [{\citenamefont {Li}\ \emph {et~al.}(2021)\citenamefont {Li}, \citenamefont {Li}, \citenamefont {Hao}, \citenamefont {Guo}, \citenamefont {Zhuang}, \citenamefont {Cui}, \citenamefont {Wei}, \citenamefont {Ma}, \citenamefont {Wang}, \citenamefont {Xu}, \citenamefont {Niu},\ and\ \citenamefont {Wang}}]{RN415}%
  \BibitemOpen
  \bibfield  {author} {\bibinfo {author} {\bibfnamefont {Y.}~\bibnamefont {Li}}, \bibinfo {author} {\bibfnamefont {X.-M.}\ \bibnamefont {Li}}, \bibinfo {author} {\bibfnamefont {R.-T.}\ \bibnamefont {Hao}}, \bibinfo {author} {\bibfnamefont {J.}~\bibnamefont {Guo}}, \bibinfo {author} {\bibfnamefont {Y.}~\bibnamefont {Zhuang}}, \bibinfo {author} {\bibfnamefont {S.-N.}\ \bibnamefont {Cui}}, \bibinfo {author} {\bibfnamefont {G.-S.}\ \bibnamefont {Wei}}, \bibinfo {author} {\bibfnamefont {X.-L.}\ \bibnamefont {Ma}}, \bibinfo {author} {\bibfnamefont {G.-W.}\ \bibnamefont {Wang}}, \bibinfo {author} {\bibfnamefont {Y.-Q.}\ \bibnamefont {Xu}}, \bibinfo {author} {\bibfnamefont {Z.-C.}\ \bibnamefont {Niu}}, \ and\ \bibinfo {author} {\bibfnamefont {Y.}~\bibnamefont {Wang}},\ }\bibfield  {title} {\enquote {\bibinfo {title} {Growth of high quality insb thin films on gaas substrates by molecular beam epitaxy method with alinsb/gasb as compound buffer layers*},}\ }\href {\doibase 10.1088/1674-1056/abc152} {\bibfield  {journal}
  {\bibinfo  {journal} {Chinese Physics B}\ }\textbf {\bibinfo {volume} {30}},\ \bibinfo {pages} {028504} (\bibinfo {year} {2021})}\BibitemShut {NoStop}%
\bibitem [{\citenamefont {Oszwaldowski}\ \emph {et~al.}(2004)\citenamefont {Oszwaldowski}, \citenamefont {Berus}, \citenamefont {Borowska}, \citenamefont {Czajka},\ and\ \citenamefont {Zimniak}}]{RN247}%
  \BibitemOpen
  \bibfield  {author} {\bibinfo {author} {\bibfnamefont {M.}~\bibnamefont {Oszwaldowski}}, \bibinfo {author} {\bibfnamefont {T.}~\bibnamefont {Berus}}, \bibinfo {author} {\bibfnamefont {A.}~\bibnamefont {Borowska}}, \bibinfo {author} {\bibfnamefont {R.}~\bibnamefont {Czajka}}, \ and\ \bibinfo {author} {\bibfnamefont {M.}~\bibnamefont {Zimniak}},\ }\bibfield  {title} {\enquote {\bibinfo {title} {Growth of insb thin films on gaas(100) substrates by flash evaporation epitaxy},}\ }\href {\doibase 10.1002/pssc.200303943} {\bibfield  {journal} {\bibinfo  {journal} {physica status solidi (c)}\ }\textbf {\bibinfo {volume} {1}},\ \bibinfo {pages} {351--354} (\bibinfo {year} {2004})}\BibitemShut {NoStop}%
\bibitem [{\citenamefont {Li}\ \emph {et~al.}(2009)\citenamefont {Li}, \citenamefont {Qiu}, \citenamefont {Liu}, \citenamefont {Wang}, \citenamefont {Zhang},\ and\ \citenamefont {Zhao}}]{RN262}%
  \BibitemOpen
  \bibfield  {author} {\bibinfo {author} {\bibfnamefont {M.}~\bibnamefont {Li}}, \bibinfo {author} {\bibfnamefont {Y.}~\bibnamefont {Qiu}}, \bibinfo {author} {\bibfnamefont {G.}~\bibnamefont {Liu}}, \bibinfo {author} {\bibfnamefont {Y.}~\bibnamefont {Wang}}, \bibinfo {author} {\bibfnamefont {B.}~\bibnamefont {Zhang}}, \ and\ \bibinfo {author} {\bibfnamefont {L.}~\bibnamefont {Zhao}},\ }\bibfield  {title} {\enquote {\bibinfo {title} {Distribution of dislocations in gasb and insb epilayers grown on gaas (001) vicinal substrates},}\ }\href {\doibase 10.1063/1.3115450} {\bibfield  {journal} {\bibinfo  {journal} {Journal of Applied Physics}\ }\textbf {\bibinfo {volume} {105}},\ \bibinfo {pages} {094903} (\bibinfo {year} {2009})}\BibitemShut {NoStop}%
\bibitem [{\citenamefont {Kanisawa}, \citenamefont {Yamaguchi},\ and\ \citenamefont {Hirayama}(2000)}]{RN275}%
  \BibitemOpen
  \bibfield  {author} {\bibinfo {author} {\bibfnamefont {K.}~\bibnamefont {Kanisawa}}, \bibinfo {author} {\bibfnamefont {H.}~\bibnamefont {Yamaguchi}}, \ and\ \bibinfo {author} {\bibfnamefont {Y.}~\bibnamefont {Hirayama}},\ }\bibfield  {title} {\enquote {\bibinfo {title} {Two-dimensional growth of insb thin films on gaas(111)a substrates},}\ }\href {\doibase 10.1063/1.125826} {\bibfield  {journal} {\bibinfo  {journal} {Applied Physics Letters}\ }\textbf {\bibinfo {volume} {76}},\ \bibinfo {pages} {589--591} (\bibinfo {year} {2000})}\BibitemShut {NoStop}%
\bibitem [{\citenamefont {Ayers}(1994)}]{RN254}%
  \BibitemOpen
  \bibfield  {author} {\bibinfo {author} {\bibfnamefont {J.~E.}\ \bibnamefont {Ayers}},\ }\bibfield  {title} {\enquote {\bibinfo {title} {The measurement of threading dislocation densities in semiconductor crystals by x-ray diffraction},}\ }\href {\doibase http://dx.doi.org/10.1016/0022-0248(94)90727-7} {\bibfield  {journal} {\bibinfo  {journal} {Journal of Crystal Growth}\ }\textbf {\bibinfo {volume} {135}},\ \bibinfo {pages} {71--77} (\bibinfo {year} {1994})}\BibitemShut {NoStop}%
\bibitem [{\citenamefont {Ayers}, \citenamefont {Ghandhi},\ and\ \citenamefont {Schowalter}(1991)}]{RN273}%
  \BibitemOpen
  \bibfield  {author} {\bibinfo {author} {\bibfnamefont {J.~E.}\ \bibnamefont {Ayers}}, \bibinfo {author} {\bibfnamefont {S.~K.}\ \bibnamefont {Ghandhi}}, \ and\ \bibinfo {author} {\bibfnamefont {L.~J.}\ \bibnamefont {Schowalter}},\ }\bibfield  {title} {\enquote {\bibinfo {title} {Crystallographic tilting of heteroepitaxial layers},}\ }\href {\doibase 10.1016/0022-0248(91)90077-i} {\bibfield  {journal} {\bibinfo  {journal} {Journal of Crystal Growth}\ }\textbf {\bibinfo {volume} {113}},\ \bibinfo {pages} {430--440} (\bibinfo {year} {1991})}\BibitemShut {NoStop}%
\bibitem [{\citenamefont {Yarlagadda}\ \emph {et~al.}(2008)\citenamefont {Yarlagadda}, \citenamefont {Rodriguez}, \citenamefont {Li}, \citenamefont {Velampati}, \citenamefont {Ocampo}, \citenamefont {Suarez}, \citenamefont {Rago}, \citenamefont {Shah}, \citenamefont {Ayers},\ and\ \citenamefont {Jain}}]{RN255}%
  \BibitemOpen
  \bibfield  {author} {\bibinfo {author} {\bibfnamefont {B.}~\bibnamefont {Yarlagadda}}, \bibinfo {author} {\bibfnamefont {A.}~\bibnamefont {Rodriguez}}, \bibinfo {author} {\bibfnamefont {P.}~\bibnamefont {Li}}, \bibinfo {author} {\bibfnamefont {R.}~\bibnamefont {Velampati}}, \bibinfo {author} {\bibfnamefont {J.~F.}\ \bibnamefont {Ocampo}}, \bibinfo {author} {\bibfnamefont {E.~N.}\ \bibnamefont {Suarez}}, \bibinfo {author} {\bibfnamefont {P.~B.}\ \bibnamefont {Rago}}, \bibinfo {author} {\bibfnamefont {D.}~\bibnamefont {Shah}}, \bibinfo {author} {\bibfnamefont {J.~E.}\ \bibnamefont {Ayers}}, \ and\ \bibinfo {author} {\bibfnamefont {F.~C.}\ \bibnamefont {Jain}},\ }\bibfield  {title} {\enquote {\bibinfo {title} {X-ray characterization of dislocation density asymmetries in heteroepitaxial semiconductors},}\ }\href {\doibase 10.1063/1.2936078} {\bibfield  {journal} {\bibinfo  {journal} {Applied Physics Letters}\ }\textbf {\bibinfo {volume} {92}} (\bibinfo {year} {2008}),\ 10.1063/1.2936078}\BibitemShut {NoStop}%
\bibitem [{\citenamefont {Sun}\ \emph {et~al.}(2014)\citenamefont {Sun}, \citenamefont {Li}, \citenamefont {Dong}, \citenamefont {Zeng}, \citenamefont {Yu}, \citenamefont {Zhao}, \citenamefont {Zhao},\ and\ \citenamefont {Yang}}]{RN274}%
  \BibitemOpen
  \bibfield  {author} {\bibinfo {author} {\bibfnamefont {Y.}~\bibnamefont {Sun}}, \bibinfo {author} {\bibfnamefont {K.}~\bibnamefont {Li}}, \bibinfo {author} {\bibfnamefont {J.}~\bibnamefont {Dong}}, \bibinfo {author} {\bibfnamefont {X.}~\bibnamefont {Zeng}}, \bibinfo {author} {\bibfnamefont {S.}~\bibnamefont {Yu}}, \bibinfo {author} {\bibfnamefont {Y.}~\bibnamefont {Zhao}}, \bibinfo {author} {\bibfnamefont {C.}~\bibnamefont {Zhao}}, \ and\ \bibinfo {author} {\bibfnamefont {H.}~\bibnamefont {Yang}},\ }\bibfield  {title} {\enquote {\bibinfo {title} {The anisotropic distribution of dislocations and tilts in metamorphic gainas/alinas buffers grown on gaas substrates with miscut angles toward (111)a},}\ }\href {\doibase 10.1016/j.jallcom.2014.01.206} {\bibfield  {journal} {\bibinfo  {journal} {Journal of Alloys and Compounds}\ }\textbf {\bibinfo {volume} {597}},\ \bibinfo {pages} {45--49} (\bibinfo {year} {2014})}\BibitemShut {NoStop}%
\bibitem [{\citenamefont {Blaikie}(2022)}]{RN423}%
  \BibitemOpen
  \bibfield  {author} {\bibinfo {author} {\bibfnamefont {T.}~\bibnamefont {Blaikie}},\ }\emph {\bibinfo {title} {Indium Antimonide Plasmonic Nanostructures for Tunable Terahertz Sources}},\ \href {http://hdl.handle.net/10012/18626} {\bibinfo {type} {Master's thesis}},\ \bibinfo  {school} {University of Waterloo} (\bibinfo {year} {2022}),\ \bibinfo {note} {available at \url{http://hdl.handle.net/10012/18626}}\BibitemShut {NoStop}%
\bibitem [{\citenamefont {Shi}\ \emph {et~al.}(2019)\citenamefont {Shi}, \citenamefont {Bergeron}, \citenamefont {Sfigakis}, \citenamefont {Baugh},\ and\ \citenamefont {Rwasilewski}}]{RN168}%
  \BibitemOpen
  \bibfield  {author} {\bibinfo {author} {\bibfnamefont {Y.}~\bibnamefont {Shi}}, \bibinfo {author} {\bibfnamefont {E.}~\bibnamefont {Bergeron}}, \bibinfo {author} {\bibfnamefont {F.}~\bibnamefont {Sfigakis}}, \bibinfo {author} {\bibfnamefont {J.}~\bibnamefont {Baugh}}, \ and\ \bibinfo {author} {\bibfnamefont {Z.}~\bibnamefont {Rwasilewski}},\ }\bibfield  {title} {\enquote {\bibinfo {title} {Hillock-free and atomically smooth insb qws grown on gaas substrates by mbe},}\ }\href {\doibase 10.1016/j.jcrysgro.2019.02.039} {\bibfield  {journal} {\bibinfo  {journal} {Journal of Crystal Growth}\ }\textbf {\bibinfo {volume} {513}},\ \bibinfo {pages} {15--19} (\bibinfo {year} {2019})},\ \bibinfo {note} {hp6he Times Cited:1 Cited References Count:23}\BibitemShut {NoStop}%
\bibitem [{\citenamefont {Mishima}\ \emph {et~al.}(2004)\citenamefont {Mishima}, \citenamefont {Keay}, \citenamefont {Goel}, \citenamefont {Ball}, \citenamefont {Chung}, \citenamefont {Johnson},\ and\ \citenamefont {Santos}}]{RN419}%
  \BibitemOpen
  \bibfield  {author} {\bibinfo {author} {\bibfnamefont {T.~D.}\ \bibnamefont {Mishima}}, \bibinfo {author} {\bibfnamefont {J.~C.}\ \bibnamefont {Keay}}, \bibinfo {author} {\bibfnamefont {N.}~\bibnamefont {Goel}}, \bibinfo {author} {\bibfnamefont {M.~A.}\ \bibnamefont {Ball}}, \bibinfo {author} {\bibfnamefont {S.~J.}\ \bibnamefont {Chung}}, \bibinfo {author} {\bibfnamefont {M.~B.}\ \bibnamefont {Johnson}}, \ and\ \bibinfo {author} {\bibfnamefont {M.~B.}\ \bibnamefont {Santos}},\ }\bibfield  {title} {\enquote {\bibinfo {title} {Effect of structural defects on insb/alxin1-x sb quantum wells grown on gaas substrates},}\ }\href {\doibase 10.1016/j.physe.2003.08.014} {\bibfield  {journal} {\bibinfo  {journal} {Physica E: Low-dimensional Systems and Nanostructures}\ }\textbf {\bibinfo {volume} {20}},\ \bibinfo {pages} {260--263} (\bibinfo {year} {2004})}\BibitemShut {NoStop}%
\bibitem [{\citenamefont {Chung}\ \emph {et~al.}(2000)\citenamefont {Chung}, \citenamefont {Ball}, \citenamefont {Lindstrom}, \citenamefont {Johnson},\ and\ \citenamefont {Santos}}]{RN420}%
  \BibitemOpen
  \bibfield  {author} {\bibinfo {author} {\bibfnamefont {S.~J.}\ \bibnamefont {Chung}}, \bibinfo {author} {\bibfnamefont {M.~A.}\ \bibnamefont {Ball}}, \bibinfo {author} {\bibfnamefont {S.~C.}\ \bibnamefont {Lindstrom}}, \bibinfo {author} {\bibfnamefont {M.~B.}\ \bibnamefont {Johnson}}, \ and\ \bibinfo {author} {\bibfnamefont {M.~B.}\ \bibnamefont {Santos}},\ }\bibfield  {title} {\enquote {\bibinfo {title} {Improving the surface morphology of insb quantum-well structures on gaas substrates},}\ }\href {\doibase 10.1116/1.591431} {\bibfield  {journal} {\bibinfo  {journal} {Journal of Vacuum Science \& Technology B: Microelectronics and Nanometer Structures Processing, Measurement, and Phenomena}\ }\textbf {\bibinfo {volume} {18}},\ \bibinfo {pages} {1583--1585} (\bibinfo {year} {2000})}\BibitemShut {NoStop}%
\bibitem [{\citenamefont {Ball}\ \emph {et~al.}(2002)\citenamefont {Ball}, \citenamefont {Keay}, \citenamefont {Chung}, \citenamefont {Santos},\ and\ \citenamefont {Johnson}}]{RN421}%
  \BibitemOpen
  \bibfield  {author} {\bibinfo {author} {\bibfnamefont {M.~A.}\ \bibnamefont {Ball}}, \bibinfo {author} {\bibfnamefont {J.~C.}\ \bibnamefont {Keay}}, \bibinfo {author} {\bibfnamefont {S.~J.}\ \bibnamefont {Chung}}, \bibinfo {author} {\bibfnamefont {M.~B.}\ \bibnamefont {Santos}}, \ and\ \bibinfo {author} {\bibfnamefont {M.~B.}\ \bibnamefont {Johnson}},\ }\bibfield  {title} {\enquote {\bibinfo {title} {Mobility anisotropy in insb/alxin1-xsb single quantum wells},}\ }\href {\doibase 10.1063/1.1463206} {\bibfield  {journal} {\bibinfo  {journal} {Applied Physics Letters}\ }\textbf {\bibinfo {volume} {80}},\ \bibinfo {pages} {2138--2140} (\bibinfo {year} {2002})}\BibitemShut {NoStop}%
\end{thebibliography}%

\end{document}